## A New Account of Personalization and Effective Communication

Douglas A. Galbi
Senior Economist
Federal Communications Commission[2]

September 16, 2001

### *Abstract*

To contribute to understanding of information economies of daily life, this paper explores over the past millennium given names of a large number of persons. Analysts have long both condemned and praised mass media as a source of common culture, national unity, or shared symbolic experiences. Names, however, indicate a large decline in shared symbolic experience over the past two centuries, a decline that the growth of mass media does not appear to have affected significantly. Study of names also shows that action and personal relationships, along with time horizon, are central aspects of effective communication across a large population. The observed preference for personalization over the past two centuries and the importance of action and personal relationships to effective communication are aspects of information economies that are likely to have continuing significance for industry developments, economic statistics, and public policy.

---

[1] The most current version is available from http://www.galbithink.org and
http://users.erols.com/dgalbi/telpol/think.htm .
[2] The opinions and conclusions expressed in this paper are those of the author. They do not necessarily reflect the views of the Federal Communications Commission, its Commissioners, or any staff other than the author. I am grateful for numerous FCC colleagues who have shared their insights and experience with me. Author's address: dgalbi@fcc.gov; FCC, 445 12'th St. SW, Washington, DC 20554, USA.

# Contents

Broad, quantitative studies of information and knowledge economies have been primarily concerned with inputs, technology, and outputs. A pioneering study pointed to the importance of knowledge growth by identifying growth in aggregate output that growth in aggregate capital and labor inputs cannot explain.[3] Other studies, including an important US Office of Telecommunications report, have used national accounting data to estimate the value of knowledge production and the share of national output associated with information activities.[4] Studies have also estimated the number of information workers and their share in the national workforce.[5] More recently, measures of technology diffusion, such as the share of persons that have telephones, computers, and Internet connections, have played prominent roles in discussion and analysis.[6]

While measures of inputs, technology, and outputs associated with information have considerable value, they also have major weaknesses. Classifying groups of workers, types of output, or output growth residuals as being associated with information involves a data naming exercise with considerable scope for discretion.[7] The results may thus provide more evidence about the particular naming exercise than about the general nature of the economy.[8] Moreover, consistent national level data on economic inputs and outputs are difficult to construct for a long period. While a long-run historical perspective is important for understanding information economies, statistical agencies face significant challenges just in coping with the effects of recent information

---

[3] Solow (1957).

[4] Machlup (1962); Porat and Rubin (1977) (the US Office of Telecommunications report); Jussawalla, Lamberton, Karunaratne (1988). The Office of Telecommunications report noted:

> It may strike some as odd that the Office of Telecommunications, an organization concerned principally with telecommunications technology, would offer a report dealing with the whole range of information activities. The explanation lies in the need to view telecommunications in the larger context of its effects on other aspects of society....To the extent that telecommunications and its [sibling] technology, computers, are at the core of the infrastructure of the information society, their relationships with the larger society are every bit as important as their internal problems.

Foreword, p. iii.

[5] The studies cited in note 2 above also provide labor force classifications and estimates. See also Schement (1990) and Castells and Aoyama (1994).

[6] See US Dept. of Commerce (1995-2000), OECD (2000), and OECD Communications Outlook (published yearly) and OECD Information Technology Outlook (published biannually).

[7] To get a sense for the classification issues, consider some summary statistics. Porat and Rubin (1977) estimated that the primary information sector accounted for 25.1% of US GNP in 1967, while the secondary information sector accounted for 21.1% of GNP. Rubin, Huber, and Taylor (1986) p. 19, updating Machlup (1962), estimated expenditures for knowledge production as 33.3% of adjusted GNP in 1967. The US National Telecommunications and Information Administration (NTIA) states on its website that by the 21'st century, telecommunications and information-related industries will account for approximately 20% of the US economy (see http://www.ntia.doc.gov/ntiahome/ntiafacts.htm, last accessed Aug. 15, 2001).

[8] The point is not that naming is unimportant; the thrust of this paper is exactly the opposite. In terms of national accounting and macroeconomics, consider, for example, the residuals presented in Solow (1957). Now widely called "total factor productivity," they probably would not have become as important if they had been more often called, in the words of Abramovitz and David (1973) p. 438, "at best a lower-bound measure of our ignorance of the process of economic growth."

technology developments.[9]    Approaches that focus on inputs, technology, and outputs also can obscure that persons are the subjects of the information economy, and that persons thinking and communicating produce non-marketed human goods and create culture for common use.[10]

Creative empirical approaches are needed to complement widely recognized theoretical developments in the economics of information. The economics of information have shaped the way economists and others think.[11]  Information is in general imperfect and asymmetric, like a tomato selected at random from a backyard garden.  Areas in which the economics of information has thus far only made limited progress include:

> *how and how well organizations and societies absorb new information, learn, adapt their behavior, and even their structures; and how different economic and organizational designs affect the ability to create, transmit, absorb, and use knowledge and information.[12]*

These questions require study that goes well beyond price systems.  The key questions relate to dynamics of the information economy not captured in traditional models of markets.

Personal given names offer several advantages for studying an information economy.[13]  On a daily basis, for most types of information, and in much of human communications, "who" and "to whom" are key questions.  Personal names matter in normal human activity, they are a crucial aspect of personal identity and dignity, and they have deep cultural significance.  Moreover, from an operational perspective, personal names have been collected extensively and over a long period of time in the process of public administration.[14]  A given name, which forms part of a contemporary personal name, is generally given to a person shortly after birth, and given names are seldom changed.[15]  Given names thus provide a means for disciplined, quantitative study of information economies across major social, economic, and technological changes.

The work of influential analysts points to the importance of studying names.  Pierre Bourdieu has declared that the social sciences must focus on "the social operations of naming," or using one of Bourdieu's distinctive terms, naming habitus, meaning a social perspective on naming habits, an aspect of which will be measured in this paper in bits.[16] Niklas Luhmann has elaborated upon the three-in-one unity (*unitas multiplex*) of information, message and understanding, and Luhmann has explored theoretically how communication constructs social systems and shifts them among different states.[17] The distribution of name frequencies, which consistently produces a particular order as part of the communication that characterizes social life, is an important empirical example of Luhmann's theory.[18] Jürgen Habermas has discussed communicative rationality in relation to the historical emergence of the public sphere, its refeudalization, and the colonization of the lifeworld.[19] Public discussion and public opinion concerning personal names affects practical private interests, such as the ability to attract attention, get respect, or communicate status. Study of names can provide important historical evidence concerning Habermas's distinction between communicative and instrumental rationality. More generally, study of names can help one better understand the widely cited work of Habermas, as well as that of Luhmann and Bourdieu.

---

[9] See Landefeld and Fraumeni (2001) and Haltiwanger and Jarmin (2000).
[10] Much work in cultural and media studies explores these issues.  A closer relationship between such work and disciplined, statistical analysis of information economies is likely to be fruitful.
[11] Stiglitz (2000).
[12] Ibid, p. 1471.
[13] Although differing somewhat from the approach in this paper, a significant literature addresses personal given names.  Much important early scholarship concerns non-English-language names, in particular French and Hungarian names.  See, for example, Michaëlsson (1927), Kálmán (1978), and Dupâquier, Bideau, and Ducreux (1980).  Non-English names desire further study and analysis, but are not considered in this paper.  With respect to medieval English names, for studies of astonishing erudition and scholarly care, see the work of Cecily Clark; in particular, Jackson (1995) and Clark (1992a,b).  Zelinsky (1970) and Scott (1985) are important early studies of naming in the US.  Wilson (1998) considers the history of naming in western Europe, with a focus on England, France, and Italy, and some treatment of the US.
[14] In Europe extensive name data is available for about a millennium.  Most countries currently collect extensive name data through censuses and registration of births and deaths.
[15] In English-language naming, the given name is typically the first name in a person's full name.  In other languages, such as Chinese and Hungarian, given names are usually the last names in a person's full name.

To contribute to understanding of information economies of daily life, this paper explores over the past millennium given names of a large number of persons. Analysts have long both condemned and praised mass media as a source of common culture, national unity, or shared symbolic experiences.[20] Names, however, indicate a large decline in shared symbolic experience over the past two centuries, a decline that the growth of mass media does not appear to have affected significantly. Study of names also shows that action and personal relationships, along with time horizon, are central aspects of effective communication across a large population. The observed preference for personalization over the past two centuries and the importance of action and personal relationships to effective communication are aspects of information economies that are likely to have continuing significance for industry developments, economic statistics, and public policy.

## I. Analyzing Names

Choosing and communicating names have long been important actions in information economies. In Hebrew scripture, the stewardship that human beings exercise over nature is expressed in God's giving the first man power to name all living creatures, and the calling and giving of names played a key role in establishing God's special relationship with Israel.[21] The classical culture of learning recognized the importance of naming in

---

[16] Bourdieu (1991) p. 105. See Section II B of this paper.
[17] See, for example Luhmann (1989), which considers communication about ecological problems.
[18] See Section I C of this paper.
[19] Habermas (1962/1989).
[20] Mass media is typically understood as mass circulation magazines and newspapers, and radio and television. Ong (1967) p. 291 suggests that mass media began with mass language. This paper does not debate the conventional definition of mass media; instead, it focuses on historical facts about shared symbolic experience and communication.
[21] *Genesis* 2:19, *Exodus* 3: 4, 13-15. In the Qur'an, naming has similar importance. See Surat 2, al-Baqarah 31-33, 163.

the Latin saying "Nomen est numen": to name is to know.  In Tudor and Stuart England (1485-1714):

> *Naming was a serious business, securing legal, social, religious, and semantic identity.  According to conventional commentators, the name given at baptism was indeed one's Christian name, a sign of 'our regeneration' and 'a badge that we belong to God'.  It also put one in fellowship with all others who had worn the name before, to be 'recorded not only in the church's register, but in the book of life, and stand there forever'.[22]*

The importance attached to naming is not anachronistic today.  The popularity of name-your-baby books and websites emphasizes that fact.[23]  Consider as well the Society for Creative Anachronism (SCA), a worldwide group of person that study and re-create the European Middle Ages.  In its activities, the SCA puts considerable emphasizes on naming.  Each SCA member adopts a unique name appropriate to the Middle Ages through a formal SCA process of authentication and registration, and in all SCA activities and communications SCA members use these names.[24]

Over the last several decades, choosing names for businesses and products has developed as a special line of commerce.  Firms such as Landor, Interbrand, Enterprise IG, Idiom, NameLab, TrueNames, and others provide commercial naming services:

> *Each of the firms has its own jealously guarded methodology, a signature "naming module" that distinguishes it from its competitors.  Enterprise IG has its proprietary NameMaker program, good for generating thousands of names by computer.  Landor uses a double-barrelled approach; deploying both its "Brand Alignment Process" and a "BrandAsset Valuator."  Others find that their module must be described in more than a few words.  "We have a wonderful approach," says Rick Bragdon of Idiom.  "We use an imaginative series of turbo-charged naming exercises, including Blind Man's Brilliance, Imagineering, Synonym Explosion and Leap of Faith…We find that when clients are playing, literally playing creative games, they create names that come from a place of joy, a place of fun.[25]*

The commercial goal is to find a "good name": a name that sounds well, that is memorable, and that has appealing connotations with respect to the particular naming situation.

As for commercial names, the value of personal names depends on norms, memories, connotations, and other aspects of shared experiences.  Norms governing naming, such as naming after parents, grandparents, biblical figures, or deceased siblings, are common laws in the economy of names. They evolve through common awareness of patterns of cases and possibilities for differences and exceptions. Estimating the value of a particular name involves collecting and assessing information about other persons' perceptions of the name within the information economy. While norms and social values structure naming choices, the actual personal choice has largely been a domain of freedom, i.e. personal preference is the recognized ultimate authority.[26] Thus chosen names provide evidence about the preferences that free persons express in a particular historical context.

---

[22] Cressy (1997) pp. 161-2.

[23] See, for example, http://www.baby-names-tips.com, http://www.namechooser.com/baby/, http://www.kabalarians.com/gkh/yourbaby.htm, and http://www.thebabiesplanet.com/bbnames.shtml for many others.

[24] The SCA, established in 1966, has more than 24,000 paying members around the world, and perhaps three or four times as many active participants (paid membership is not required for participation in SCA activities).  For general information on the SCA, see http://www.sca.org/sca-intro.html  The SCA has a College of Arms that reviews, authenticates, and registers members' choices of names.  For details on this process, see the SCA, Rules of Submissions of the College of Arms, Parts I through VI [online at http://www.sca.org/heraldry/laurel/regs.html ].

[25] Shalit (1999) p. 4.

Personal given names relate to a significant part of shared symbolic experience. Persons who have the same given name literally share the experience of being called by that name; they share the experience of being associated with all the social meaning attached to the name. Birth parents and chosen others, such as godparents, also share the experience of determining a good name for another person. Through the course of their lives persons have a wide range of other symbolic experiences. Naming, however, is probably unique in its combination of personal significance, universal prevalence, and consistency through time.

## A. Charting Name Trends

How to analyze given names and their changes over time is not obvious. One might ponder why particular names are chosen and think about factors that affect popularity trends. A recent book, entitled *A Matter of Taste,* sought to develop theory to address such issues.[27] A chapter entitled, "Broader Issues: The Cultural Surface and Cultural Change," moves from subsections labeled "A Causal Hierarchy" and "Birth and Death Are Not the Same"[28] to one labeled "Monica."[28] This subsection used the theory developed in the book to consider how the sexual liaison between President Bill Clinton and Monica Lewinsky would affect the popularity of the name Monica.

The author's analysis is interesting. First, he notes that "the necessary basis for making a prediction is more complicated that it might appear."[29] He then commences by visualizing four possibilities: 1) the name was rarely used, 2) the name was gaining in popularity in the preceding years, 3) the name was failing in popularity in the preceding years, 4) the name was relatively stable in popularity. Without the Lewinsky affair, "In each case, our best expectation would be more of the same," although "in some small proportion of the time we would be wrong."[30]

---

[26] In the economics literature such freedom is generally referred to as "consumer sovereignty."
[27] Lieberson (2000). Lieberson is Abbott Lawrence Lowell Professor of Sociology at Harvard University.
[28] Ibid, pp. 261-6.
[29] Ibid, p. 263. Lieberson is quite modest in the claims he makes for his ideas. He states: "These are ideas about what may play a role in driving tastes. [footnote omitted] They may be helpful in any given context. If they are, that's fine, and we are happy to use them. If they aren't, it doesn't mean that the ideas are worthless or wrong. Rather, they don't seem to work in the context of the particular conglomeration of historical and external and internal conditions. That is all one can say." Ibid, p. 21.
[30] Ibid, p. 263-4.

The analysis of "what can cause something to *happen* that differs from these expectations" is essentially the same in all four cases.[31]  Here's the analysis:

> *We could say that a modest proportion of parents,* m*, had been using the name* Monica *and that a far larger proportion of parents,* o*, had been using some other name.  If we can assume that the new set of parents in the year following the scandal had identical dispositions, then the net movement of* Monica *is the product of two transitions: what number of the* m *population are now turned away from the name and what number of the* o *population are now turned toward the name.  The difference between the two will mean that* Monica *gains or loses in popularity.  Again, because we start with so many more people initially disposed not to use* Monica*, it takes only a small proportion of the* o *parents to switch for the name to gain in popularity even if the vast number of* m *are no longer attracted to the name.*[32]

The author does not provide any specific prediction about changes in the popularity of Monica.  He does, however, note "how easy it is to misinterpret the eventual answer – no matter what that answer is."[33]

The above analysis, and much of the rest of the analysis in *A Matter of Taste*, is similar to what is known in the financial world as technical analysis.  Technical analysis concerns the study and interpretation of stock price trends separate from external factors or the fundamental value of a company.[34]  The focus is on "internal mechanisms" that drive price movements, such as momentum and symbolic enhancement or contamination from crossing levels of support or resistance (usually round numbers like multiples of ten or a hundred).[35]  Such analysis is commonplace in the financial world and a regular part of mainstream financial reporting.[36]

---

[31] Ibid, p. 264.

[32] Ibid.

[33] Ibid, p. 263.

[34] Charles Dow, a journalist who was the first editor of the *Wall Street Journal* and one of the founders of the Dow Jones & Company, was an early, influential practitioner of this sort of analysis.  Under Dow, the Wall Street Journal published the first stock index, an average of 12 active stocks.  This index evolved into the now widely quoted US stock index, the Dow Jones Industrials Average, usually just called the Dow.  Dow wrote a large number of editorials and columns, and his views are not easily summarized (see, for example, the bland, incomplete list of twenty four observations or principles put forth in Bishop (1967) pp. 305-6).  Dow stated, "Nothing is more certain than that the market has three well-defined movements which fit into each other."  These movements Dow categorized as daily variation, secondary movement (10-60 day cycle, averages 30-40 days) and great swing (4-6 years).  He used such a conceptual scheme to discuss movements in stock indices.  See Bishop (1960) p. 56.  Dow's friend S.A. Nelson attempted to explain Dow's analysis in a book, *The ABC of Stock Speculation* (1903).  For recent expositions of technical analysis, see, for example, the descriptive material online at TAguru.com, at http://taguru.com/educational.html and Equity Analytics, at http://www.e-analytics.com/techdir.htm , and Steven B. Achelis's online textbook, *Technical Analysis from A to Z*, at http://www.equis.com/free/taaz/index.html

[35] Cf. Lieberson, pp. 93-98, 126-42.

[36] See, for example, "Stevens on Technical Analysis," a weekly column on CNBC [http://www.cnbc.com/010712stevens-stocks.html ].  For a skeptical view of technical analysis, see Mann, Bill, "Is Technical Analysis Voodoo?" on The Motley Fool [online at http://www.fool.com/news/foth/2001/foth010105.htm ].

While technical analysis provides a rich discourse for discussing observed trends and possible future developments, this paper seeks effective tools for uncovering hidden truths about information economies. Three scientific virtues will guide the analysis: observability, simplicity, and consistency. Important factors that affect the popularity of particular names may be difficult to observe, they may be many and complex, and they may vary significantly across names. Thus the analysis will not address the popularity of particular names. Instead, it will focus on characteristics of the over-all sample of names, characteristics that can be informatively measured in actual name samples of about 500 or more English-language names.

## B. Statistically Measuring Names

Names present some subtle statistical challenges. A sample of persons' names may cover a significant share of the finite population under study. Thus statistical issues associated with finite samples are relevant. Moreover, the abstract sample space of names is of very high dimension, and all samples sparsely populate that space. Thus the natural space of names as tokens is awkward to manipulate. One way to simplify the sample space is to define the sample as a token frequency distribution. A disadvantage is that the sample space then becomes a function of the sample size. In such a context analysis of properties of estimators is complex.

Rather than exploring such statistical issues abstractly, this paper takes an operational approach. Conditional on interest in a particular name or set of names, the distribution of names is a binomial or multinomial distribution. Based on available name sample sizes and associated sampling errors, the desirability of powerful statistics, and empirical evaluation of alternatives, this paper focuses on the ten most popular names in a sample.[37] The values of the statistics in this paper depend on the rank cutoff used in the analysis. However, the overall trends observed do not appear to depend on this choice.[38]

Measuring name frequencies in actual samples requires attention to name definition and standardization. Given names can include multiple names and name variants, as well as abbreviations, non-standard spellings, and mistakes in recording. Throughout the analysis in this paper, names have been truncated to the shorter of either the first eight letters of the given name or the letters preceding the first period, space, hyphen, or other non-alphabetic character. These shortened names have then been standardized through a name coding available on the Internet for public inspection, use, and improvement on an open source basis.[39]  This procedure attempts to identify feasibly and consistently names with common communicative properties.[40]

---

[37] Focusing on high frequency names also has the benefit of limiting the number of standardization choices that significantly affect the analysis. See subsequent paragraph in text.
[38] Based on analysis of US and UK name samples for the nineteenth and twentieth centuries. Note that name popularities are highly regular; see the subsequent section, especially Chart 1.

For name samples comprising between 1,000 and 10,000 names, coding inconsistencies appear to be similar in magnitude to sampling variability.   Table 1 shows sampling variability for a single name, given different probabilities for the name in the population and different sample sizes.  Sampling variability is likely to be insignificant in modern name samples that can easily comprise over a million names.  Medieval name samples, however, are often limited to 1000 names or less.  For such sample sizes, sampling variability can easily account for a percentage point difference in a name frequency statistic.  The importance of coding depends on the particular name, time, place, and recording process.  Table 2 shows name variants coded to "Mary" for UK and US name samples in different periods.   Clearly coding matters, but the nature of coding errors and inconsistencies is more speculative.  Experience with different name samples from the same population suggests that coding variability can be reduced to less that half a percentage point for the frequency of a single name and less than three percentage points for total frequency of the top ten names.[41]

| Table 1 Sampling Variability for Name Popularity | | | | |
|---|---|---|---|---|
| Name Probability | Sample Size | Expected Name Freq. | Standard Deviation | Std. Dev. (% of sample) |
| 20.0% | 100 | 20 | 4 | 4.0% |
| 3.0% | 100 | 3 | 2 | 1.7% |
| 1.5% | 100 | 2 | 1 | 1.2% |
| 20.0% | 1,000 | 200 | 13 | 1.3% |
| 3.0% | 1,000 | 30 | 5 | 0.5% |
| 1.5% | 1,000 | 15 | 4 | 0.4% |
| 20.0% | 10,000 | 2,000 | 40 | 0.4% |
| 3.0% | 10,000 | 300 | 17 | 0.2% |
| 1.5% | 10,000 | 150 | 12 | 0.1% |
| 20.0% | 100,000 | 20,000 | 126 | 0.1% |
| 3.0% | 100,000 | 3,000 | 54 | 0.1% |
| 1.5% | 100,000 | 1,500 | 38 | 0.0% |

[39] See the GINAP site, http://users.erols.com/dgalbi/names/ginap.htm.  The principle for coding is to group together names that either sound the same, have the same public meaning, or changed only in the recording process (spelling errors, recording errors, etc.).

[40] Note that name standardization helps to control for changes in names used as a person grows older, e.g. a correlation between nicknames or informal names and age.  Thus name standardization is particularly important in analyzing time trends when the data come from naming cohorts constructed by age.  That is the case for this paper's data on nineteenth century names.

[41] These estimates refer to coding variability after name standardization using GINAP, version 1.  GINAP is available at http://users.erols.com/dgalbi/names/ginap.htm.  For further data and discussion of variations in name statistics, see Appendix B.

| Table 2 Names Coded to Mary | | | | | |
|---|---|---|---|---|---|
| US | | | UK | | |
| Years | Name | Popularity | Year | Name | Popularity |
| 1810-1819 | Mary | 7.6% | 1820 | Mary | 18.1% |
| | Mary A | 1.8% | | Maria | 1.9% |
| | Maria | 1.1% | | Maryann | 0.1% |
| 1900-1910 | Mary | 5.6% | 1900 | Mary | 3.9% |
| | Marie | 1.3% | | Marion | 0.3% |
| | Marion | 0.6% | | Maria | 0.3% |
| 1990-1999 | Mary | 0.5% | 1975 | Marie | 0.6% |
| | Maria | 0.5% | | Maria | 0.2% |
| | Marissa | 0.3% | | Mary | 0.1% |

Note: For sources for all the name statistics in this paper, see Appendix D and References.

## C. An Important Empirical Regularity

For names occurring sufficiently frequently, name frequencies follow a power law. This means that, to a good approximation, name frequency is log-linearly related to frequency rank. Chart 1 shows on logarithmic scales the relationship between name frequency and frequency rank for females born in the US in 1831-40 and in 1990-99. While some concavity is evident, in each case a line provides a high goodness of fit.[42]

---

[42] The R-squared statistics for a log-linear least square fit for the data for 1831-40 and 1990-99 are 0.95 and 0.88, respectively.

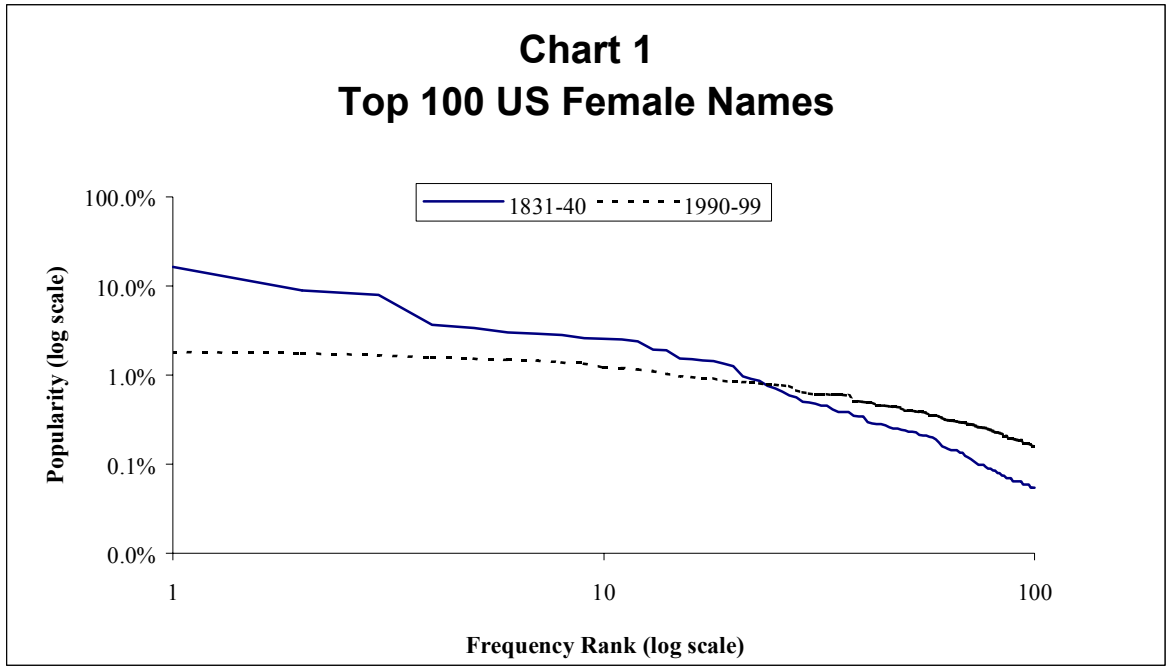

**Chart 1**
**Top 100 US Female Names**

This empirical regularity is important for several reasons. First, it highlights an order associated with naming that is potentially amenable to explanation.[43] Second, it provides a basis for describing changes in naming over time. As Chart 1 shows, the slope and position of the line describing the relationship between log-rank and log-frequency has changed significantly from 1831-40 to 1990-99. Changes in these parameters have taken a relatively smooth path that can be summarized simply. Third, a variety of other phenomena, such as word frequencies, city sizes, income distribution, and the proportion of rock surface area that barnacles, mussels and other organisms occupy in an intertidal zone, follow power laws.[44] Through this common regularity, evidence and insights regarding other phenomena can be related to naming, and insights from the study of naming gain more general significance.

Power laws are in fact prevalent in the information economy. Where persons and organizations are free to create and choose among many collections of symbols instantiated and used in a similar way, the relative popularity of the symbolic artifacts typically follows a power law.[45] Thus the circulation of magazines of similar type have followed power laws throughout the twentieth century.[46] The total box office receipts of movies follow a power law.[47] The popularity of musical groups, as measured by "gold records," follows a power law.[48] The popularity of Internet web sites, measured in users or page views, also follows a power law.[49] Insights into the evolution of such power laws over time can provide insights into personal preferences, media diversity, and industry structure in the information economy.

---

[43] An example might help to illustrate the point here. It would be very difficult to explain why the outcome of a particular flip of a coin came up heads or tails. But if one notices that the proportion of heads across a large number of flips is one-half, one might develop a statistical explanation of coin-flipping that offers considerable insight into outcomes.

[44] See Simon (1955) and Gell-Mann (1994) pp. 92-100, 319-20. Power laws are associated with random growth processes of a proportional form. If the mean and variance of the growth process are independent of the magnitude of the data under consideration and the number of data points is fixed, the relationship between log-rank and log-size (frequency) will converge to a linear relation with slope equal to −1. This relationship with respect to city sizes is known as Zipf's Law. For an insightful discussion of Zipf's Law and proof of the above result, see Gabaix (1999). Gabaix provides also provides interesting analysis of the effect of birth and death of the elements subject to the growth process. With respect to names, coding errors and classification ambiguities make counting new names subject to considerable uncertainty; however, preliminary analysis suggests that in the US from 1831-40 to 1901-10 the number of distinct names in use grew faster than the size of the population using the names.

## II. A Statistical History of Personalization

Mass media create shared symbolic experiences by producing and distributing common packages of symbols to large numbers of persons. As little as one and half centuries ago, sharing symbols was largely a matter of decentralized, peer-to-peer diffusion, performances, public meetings, monuments, and other special-purpose artifacts. In contrast, in many countries today, through mass media millions of persons regularly experience exactly the same presentations of sports, news, songs, and dramatic stories.

Concern about the role of mass media in shaping shared experience has been commonplace. As early as the mid-1940s observers warned that applying industrial technology and organization to symbol production and distribution was producing a "ruthless unity," "the same stamp on everything," a world in which "[t]he might of industrial society is lodged in men's minds," and "[r]eal life is becoming indistinguishable from the movies."[50] By the early 1990s, the same assumptions about the facts prevailed, but a sense of nostalgia had developed, at least among some:

> For forty years we were one nation indivisible, under television. That's ending. Television is turning into something else, and so are we. We're different. We're splintered. We're not as much 'we' as the 'we' we were. We're divisible.[51]

Many policy analysts and policy makers have considered mass media necessary both to promote diversity and to encourage national unity, and they have balanced according to current needs these important social and cultural values.[52]

---

[45] Vogel (2001) pp. 352-4 provides a useful list of frequently observed characteristics in the entertainment industry. Power laws might be considered an additional such characteristic.

[46] Galbi, Douglas, unpublished analysis.

[47] Galbi, Douglas, unpublished analysis.

[48] Chung and Cox (1994) .

[49] Adamic and Huberman (2000), Alexa Research (2000). See also the interview with Brewster Kahle of Alexa at http://www.feedmag.com/re/re392.2.html

[50] Horkheimer and Adorno (1944/1972) pp. 123, 120, 127, 126.

[51] Shales (1991), cited in Webster and Phalen (1997), citing Dizard (1994).

[52] In the US, which has considerable cultural diversity and a strong individualistic ethic, media policy has focused on diversity rather than unity. For discussion of the FCC's diversity analysis, see FCC (1995). The balance in other countries is different. According to an account of an OECD discussion of broadcasting regulation, "The United Kingdom added that the debate in Brussels in recent months has shown that there is still a political desire to see national markets delivering national identities, especially in smaller countries. For example, the responses to the convergence Green Paper showed a focus on content delivering national cultural identities." OECD (1999) p. 121.

New computing and communications technologies may significantly affect the extent of shared experiences. For most persons, purchasing goods and services is a significant shared experience; in the US, retail chains such as Wal-Mart, CVS, and 7-Eleven are real icons of consumer life.[53] Some have argued that e-commerce and associated personalization technologies will radically reshape retailing.[54] Particularly in societies in which a common experience of continually increasing material prosperity is an important political ideal, this change in shared experiences might present risks of political fragmentation and polarization. New technologies are also expanding opportunities to personalize education, entertainment, news, and other forms of digital content.[55] These new opportunities might lead to a reduction in shared symbolic experiences, less exposure to diverse views and topics, more social fragmentation, and more group polarization.[56]

Discriminating between the possible and the likely is worth attempting. Given the vast opportunities for personalization in the information economy, a fundamental issue is whether opportunities for personal choices will lead to similar choices or diverse choices. Similar choices might be produced from common, primal attractions such as sex, violence, and truth, from bandwagon, fashion, or tipping effects, or from social structures and institutions that homogenizes habits and preferences. Diverse choices might express the uniqueness of each person, the requirements and processes of innovation and creativity, or social forces promoting differentiation and individualism. Carefully interpreted facts with respect to aggregate symbolic choices could offer important insights into the potential social and economic significance of expanding technological possibilities for symbolic personalization.

## A. Changes in Name Popularity

The popularity of the most popular given name provides an informative indicator of shared symbolic experiences. In both the UK and US early in the nineteenth century, the most popular names were highly popular. Table 3 shows that in the UK in 1800, 23.9% of females were named Mary, the most common female name. Since living siblings almost never bore the same given name and the average fecund marriage produced 3.28 recognized daughters, the share of married women who had a daughter named Mary was

---

higher, probably about 30%.[57]   This represents a high degree of social consensus about an important symbol.

That the name Mary would generate such consensus is particularly remarkable given the bitter split between the Church of England and the Roman Catholic Church.  Roman Catholicism highly venerates Mary, the mother of Jesus.  Anti-Catholicism in England since the mid-sixteenth century has included contempt for Catholic veneration of Mary. Catholics were associated with irrational and idolatrous religious representation in which the name Mary figured highly:

> *A Papist is an Idolater, who worships Images, Pictures, Stocks and Stones, the Works of Men's Hands; calls upon the Virgin* Mary *[distinctive typeface in original], Saints and Angels to pray for them…*[58]

Yet, judging from names, there must have been something about Mary for ordinary English persons early in the nineteenth century.[59]

---

[57] Note that any daughters born subsequent to a (living) daughter named Mary largely were not at risk of being called Mary in early nineteenth century England.  (The situation was different from the twelfth to the sixteenth century.  See Wilson (1998) pp. 231-3.)  Looking at the share of women with a daughter named Mary controls for the norm that names are not repeated among living siblings.  Such a statistic is important for long-term comparisons because family sizes have changed dramatically over the past two centuries. The number of recognized (legitimate) daughters produced in a fecund marriage is calculated from data in Livi-Bacci (2000) Table 5.1, p. 95.  The proportion of births to unmarried couples was about 5% in England circa 1800 (Laslett, Costerveen, and Smith (1980) p. 14). Calculation assumes that half the daughters named Mary were first-born daughters, and half were second-born.  Calculation adjusts for daughters produced by unmarried couples and children's death rates.  Women born in 1954 in England and Wales who had children had on average 0.82 daughters.  See ONS (1999) Tables 10.4 and 10.5.

[58] *The Weekly Observer*, June 1716, quoted in Haydon (1994) p. 22.

[59] For further data and analysis, see Appendix C.

| Table 3 Most Popular Names in the UK | | | | | | | |
|---|---|---|---|---|---|---|---|
| | Females | | | Males | | | |
| Birth Year | Top Name | Pop. | Top 10 Pop. | Top 10 Info $I_s$ | Top Name | Pop. | Top 10 Pop. | Top 10 Info $I_s$ |
| 1800 | Mary | 23.9% | 82.0% | 0.511 | John | 21.5% | 84.7% | 0.356 |
| 1810 | Mary | 22.2% | 79.4% | 0.465 | John | 19.0% | 81.4% | 0.299 |
| 1820 | Mary | 20.4% | 76.5% | 0.433 | John | 17.8% | 80.4% | 0.274 |
| 1830 | Mary | 19.6% | 75.8% | 0.372 | John | 16.4% | 78.2% | 0.244 |
| 1840 | Mary | 18.7% | 75.0% | 0.333 | William | 15.4% | 76.0% | 0.231 |
| 1850 | Mary | 18.0% | 72.1% | 0.315 | William | 15.2% | 73.8% | 0.220 |
| 1860 | Mary | 16.3% | 68.3% | 0.265 | William | 14.5% | 69.8% | 0.209 |
| 1870 | Mary | 13.3% | 61.1% | 0.193 | William | 13.1% | 63.5% | 0.173 |
| 1880 | Mary | 10.6% | 53.8% | 0.116 | William | 11.7% | 58.9% | 0.144 |
| 1900 | Elizabet | 7.2% | 38.5% | 0.079 | William | 9.0% | 50.9% | 0.086 |
| 1925 | Mary | 6.7% | 38.7% | 0.070 | John | 7.3% | 38.0% | 0.100 |
| 1944 | Margaret | 4.5% | 31.7% | 0.050 | John | 8.3% | 39.9% | 0.181 |
| 1954 | Susan | 6.1% | 32.5% | 0.078 | David | 6.3% | 37.8% | 0.112 |
| 1964 | Susan | 3.6% | 28.6% | 0.022 | Paul | 5.6% | 39.4% | 0.073 |
| 1974 | Sarah | 4.9% | 28.0% | 0.089 | Mark | 4.6% | 33.1% | 0.033 |
| 1984 | Sarah | 4.1% | 27.3% | 0.049 | James | 4.3% | 32.3% | 0.021 |
| 1994 | Emily | 3.4% | 23.8% | 0.023 | James | 4.2% | 28.4% | 0.035 |

Note: See Appendix D and References for sources.

The position of Mary in the UK exemplifies the high popularity of the most popular names in other English-language populations early in the nineteenth century. In the UK, 21.5% of males born in 1800 were named John, the most popular male name. In the US, 15.0% and 12.7% of females and males born in 1800-1809 were named Mary and John, respectively, which were the most popular female and male names in those years (see Table 4). The differences between the UK and the US, while deserving further study, might plausibly be related to their much different patterns of settlement and group formation.

Over the past two centuries, the most popular names in both the UK and the US have become much less popular. In the UK from 1800 to 1994, the popularity of the most popular female and male names fell from 23.9% and 21.5% to 3.4% and 4.2%, respectively. In the US, the popularity of the most popular female and male names declined from 15.0% and 12.7% to 2.2% and 2.7%. Given the sharp reduction in average family size over the past two centuries, the extent of consensus in feasible naming choices fell even more than these simple statistics indicate.[60]

---

[60] See previous text and footnote in this section.

Moreover, the total popularity of the ten most popular names, a figure much greater than the popularity of the most popular name, also fell sharply over the past two hundred years. As Tables 3 and 4 show, in the UK and US the share of persons bearing the ten most popular names fell by roughly three times or more from about 1805 to about 1995. Based on the evidence of name popularities, the extent of shared symbolic experiences has decreased significantly over the past two centuries.

| | Females | | | | Males | | | |
|---|---|---|---|---|---|---|---|---|
| Year | Top Name | Pop. | Top 10 Pop. | Top 10 Info $I_s$ | Top Name | Pop. | Top 10 Pop. | Top 10 Info $I_s$ |
| 1805 | Mary | 15.0% | 53.7% | 0.333 | John | 12.7% | 46.8% | 0.262 |
| 1815 | Mary | 14.9% | 54.5% | 0.320 | John | 12.3% | 48.7% | 0.259 |
| 1825 | Mary | 15.8% | 55.3% | 0.334 | John | 12.1% | 48.5% | 0.257 |
| 1835 | Mary | 15.7% | 53.4% | 0.342 | John | 11.6% | 49.5% | 0.242 |
| 1845 | Mary | 16.1% | 50.8% | 0.346 | John | 11.5% | 50.5% | 0.232 |
| 1855 | Mary | 14.6% | 47.2% | 0.277 | John | 11.0% | 50.4% | 0.202 |
| 1865 | Mary | 12.3% | 43.2% | 0.230 | John | 10.0% | 50.3% | 0.195 |
| 1875 | Mary | 10.1% | 37.6% | 0.209 | William | 9.1% | 46.3% | 0.182 |
| 1885 | Mary | 7.6% | 32.6% | 0.171 | William | 7.3% | 40.1% | 0.136 |
| 1895 | Mary | 7.1% | 29.6% | 0.172 | William | 6.0% | 33.6% | 0.111 |
| 1905 | Mary | 6.8% | 28.4% | 0.164 | John | 5.0% | 29.0% | 0.083 |
| 1915 | Mary | 7.9% | 30.4% | 0.179 | John | 5.3% | 31.0% | 0.089 |
| 1925 | Mary | 8.6% | 30.0% | 0.211 | William | 5.7% | 34.4% | 0.130 |
| 1935 | Mary | 7.4% | 29.4% | 0.144 | Robert | 6.3% | 37.3% | 0.122 |
| 1945 | Mary | 5.9% | 30.6% | 0.081 | James | 5.8% | 37.2% | 0.093 |
| 1955 | Mary | 4.5% | 28.6% | 0.080 | Michael | 4.5% | 35.3% | 0.047 |
| 1965 | Elizabet | 4.0% | 22.4% | 0.070 | Michael | 4.7% | 32.3% | 0.055 |
| 1975 | Christin | 3.5% | 21.3% | 0.079 | Michael | 4.3% | 28.0% | 0.035 |
| 1985 | Christin | 3.2% | 20.9% | 0.042 | Michael | 3.6% | 24.6% | 0.037 |
| 1995 | Christin | 2.2% | 15.9% | 0.018 | John | 2.7% | 18.3% | 0.036 |

**Table 4**
**Most Popular Names in the US**

Note: The data refer to persons named (born) in the ten years around the year given. For details and sources, see Appendix D.

## B. An Information-Theoretic Statistic

The change in name popularities reflects a change in the shape of the name popularity distribution. As noted in Section I C above, the name popularity distribution follows a power law. The reduction in popularity of the most popular name implies a reduction in the intercept of a line approximating the log-rank log-frequency relation. The popularity of the ten most popular names relates to both the intercept and slope of the approximating line. As Chart 1 shows, the slope has become flatter over time. The effect of the change in intercept and slope is such that the ten most popular names cover a smaller share of the population, and changes in name popularities across the ten most popular names are relatively smaller.

Information theory provides an insightful alternative to power law approximations in describing the name distribution and changes in it over time. To understand the relevance of information-theoretic measures, consider the following scenario. Suppose that all females are named Mary. Then being told a female's name communicates no information. There is complete social consensus about the value of the name Mary, as revealed in actual naming choices, and all females share the experience of being called Mary. More generally, the social pattern of naming indicates a relatively high amount of common information and shared experience.

Now consider a different scenario. Suppose that all female names are equally likely. Each name may itself carry significant social meaning, perhaps such as Hilda, a traditional English name; Chastity, a virtue name associated with Puritans; or Brittany, a name with little history but recent prominence. However, one can do no better than to guess randomly about who is Chastity and who is Brittany. The popularity distribution of names, an important aspect of the social structure of naming, provides no additional information. In this sense the information associated with naming is wholly personal.

Information-theoretic statistics capturing the above considerations are well known.[61] Equation 1 defines information statistic $I_s$ in terms of name popularities $p_i$. $I_s$ represents the amount of social information associated with the popularity distribution of the most popular ten names. $I_s$ is related to the slope of the popularity distribution; more social information is associated with a steeper slope.

$$(1) \quad I_s = \log_2(n) - \sum_{i=1}^{10} \frac{p_i}{T} \log_2\left(\frac{p_i}{T}\right), \text{ where } T = \sum_{j=1}^{10} p_j$$

As a statistic, $I_s$ has the advantage of being measured in bits. Bits have a specific meaning in terms of coding information, and changes in $I_s$ represent changes in bits of social information. In contrast, the popularity of the top name and top ten names are percentages. They are dimensionless numbers, and an absolute or relative change in percentages is difficult to interpret quantitatively.

The trend in social information is similar to the trends in the popularity of the most popular name and the total popularity of the top ten names. This is not merely an arithmetic tautology. Since $I_s$ depends only on relative name popularities, the popularity of the top name and total popularity of the top ten names could fall sharply while $I_s$ remained constant. In fact, as Tables 3 and 4 show, all three have fallen dramatically. Over the past two hundred years, $I_s$ has fallen by about a factor of ten, with the reduction for females being roughly three times as great as that for males.

---

[61] For an introduction to information theory from an economic perspective, see Theil (1967). By presenting non-information theoretic statistics and a considerable amount of evidence interpreted without any regard to information theory (see previous Section II A), this paper hopes to avoid being grouped with a series of papers generically entitled "Information Theory, Photosynthesis, and Religion." See Elias (1958).

Changes in $I_s$ over the past two hundred years track the gross shape of changes in the popularity of the top name and the top ten names. For both the UK and the US, all three series have trendless periods of 25-50 years at some point late in the nineteenth to early in twentieth centuries, with the trendless period for $I_s$ coming slightly (10-20 years) earlier than the trendless period for the popularity statistics (see Tables 3 and 4). For the US, all three series show change concentrated in the mid-nineteenth century and in the second half of the twentieth century. The UK has a roughly similar pattern in the twentieth century, but shows a downward trend throughout the nineteenth century. The similarities between changes in $I_s$ and changes in the name popularity statistics suggests that all these statistics are measuring the same underlying change in the information economy, a change which this paper will call an increase in personalization.

## C. More than a Thousand Years of Information Economy History

Additional historical evidence helps to provide insight into name personalization and the information economy. In early medieval England, personal names consisted of a single word, typically formed from a combination of two elements associated with name-words. A large number of different personal names could thus be constructed.[62] The repetition of names among persons related through blood, time, or space could hinder identification or violate the order of the spiritual world. Repetition of names was not a general practice.[63] The extensive name personalization that characterizes the late twentieth century US and UK appears to have been also a feature of early medieval England before the tenth century.

The disproportionate favoring of a few names seems to have emerged in England during the tenth and eleventh centuries.[64] Edmund, King of the East Angles late in the ninth century, was widely admired. Moreover, he was martyred by the Danes. He may have played an important part in personalizing the position of king and inspiring widespread repetition of his particular name.[65] The change in the social pattern of names suggests the development of a public sphere in which the value of particular names, and the merits of the king as a person like other persons, were subject to discussion.[66] Significant economic development also probably occurred in the tenth and eleventh centuries. By 1066 more than 30% of economic output was marketed, and three-quarters of that entered international trade.[67] The rise in social information and shared experience in naming thus occurred along with personalized celebrities and a significant volume of commercial transactions.

---

[62] Clark (1992a), pp. 456-9.
[63] Withycombe (1945/1977) p. xxiv. Sharing of elements of names or alliteration were used at least in some cases to mark kinship. See Clark (1992a) p. 458.
[64] Stenton (1924/1970) pp. 94-6, Clark (1992a) p. 461, Clark (1992b) pp. 552, 555.
[65] Seltén (1972) p. 36.
[66] Cf. Habermas (1962/1989) pp. 7-9.
[67] Snooks (1995) p. 27.

The Norman Conquest of England in 1066 produced a dramatic change in given names. Within a few generations, most persons used given names brought by the invaders. By about 1250 pre-Conquest names had essentially died out.[68] The influx of new names and the shift to them must have decreased the popularity of the most popular names until the new naming practices were well established throughout society. Thus an increased social flow of information, specifically, the social transmission of a new set of names, can coincide with a reduction in the level of social information and shared experience in naming.

The available evidence indicates that more popular names increased in popularity from about the beginning of the twelfth century through the beginning of the fifteenth century. Table 5 provides statistics for different areas and from different sources.[69] As the shift away from pre-Conquest names progressed, Norman names became the most popular names, and the popularity of the most popular names increased.[70] The Black Death in 1347-49 and associated economic hardships appear to have prompted a sharp, further increase in the popularity of the most popular names.[71] Such a change might indicate social solidarity in reaction to the catastrophe, at least with respect to social information and shared experience in naming.

---

[68] Clark (1992b) pp. 552, 558-562. There is no evidence that Norman clergy or royal officials compelled the English to adopt Norman names.

[69] While further collection and analysis of primary sources would be useful, the rough outlines of historical change seem clear.

[70] With respect to the names in Table 5, note that Robert, William, Alice, Matilda, and John were not popular names in the pre-Conquest period. See Barlow et. al. (1976) Table 8, p. 187.

[71] See Table 5. In contrast to 1350, Manchester, see 1350, Yorkshire; 1350, North/Cumbria; 1350, Hereford; 1385, soldiers; and 1400, East Anglia.

## Table 5
## Name Popularity in England before c. 1825

| Year, Location | Females | | | | | Males | | | | |
|---|---|---|---|---|---|---|---|---|---|---|
| | Top Name | Pop. | Top 10 Pop. | Top 10 Info $I_s$ | Sample Size | Top Name | Pop. | Top 10 Pop. | Top 10 Info $I_s$ | Sample Size |
| 1080, Winchester 1) | | | | | | Robert | 6.6% | 35% | 0.16 | 228 |
| 1120, Winchester 1) | | | | | | William | 6.6% | 30% | 0.24 | 912 |
| 1180, Winchester 1) | | | | | | William | 10.2% | 57% | 0.15 | 383 |
| 1200, Essex 2) | Alice | 11.3% | 56% | 0.17 | c. 1400 | William | 12.4% | 61% | 0.18 | c. 4000 |
| 1210, South 3) | Matilda | 16.2% | 70% | 0.27 | 173 | William | 14.4% | 65% | 0.16 | 877 |
| 1270, Rutland 4) | Alice | 19.4% | 84% | 0.32 | 206 | William | 15.2% | 76% | 0.16 | 1627 |
| 1300, Lincoln 5) | Alice | 17.1% | 75% | 0.32 | 1213 | John | 22.7% | 79% | 0.47 | 9390 |
| 1260, London 6) | | | | | | John | 17.6% | 69% | 0.32 | 814 |
| 1290, London 6) | | | | | | John | 23.3% | 73% | 0.42 | 1852 |
| 1510, London 7) | | | | | | John | 24.4% | 74% | 0.53 | 427 |
| 1610, London 7) | | | | | | John | 21.0% | 72% | 0.36 | 463 |
| 1825, London 8) | Mary | 19.2% | 73% | 0.38 | 63809 | William | 16.3% | 80% | 0.19 | 48275 |
| 1350, Manchester 9c) | | | | | | John | 22.7% | 92% | 0.32 | 717 |
| 1610, Manchester 10) | | | | | | John | 18.6% | 77% | 0.21 | 1298 |
| 1805, Manchester 11) | Mary | 25.8% | 84% | 0.34 | 1866 | John | 21.7% | 81% | 0.39 | 1935 |
| 1350, Yorkshire 9d) | Alice | 22.4% | 86% | 0.30 | 1794 | John | 33.5% | 93% | 0.72 | 1665 |
| 1620, Yorkshire 12) | Ann | 24.0% | 88% | 0.47 | 342 | John | 16.2% | 86% | 0.21 | 427 |
| 1670, Yorkshire 12) | Ann | 21.5% | 79% | 0.69 | 228 | William | 18.7% | 78% | 0.34 | 283 |
| 1720, Yorkshire 12) | Mary | 25.7% | 87% | 0.57 | 413 | John | 25.5% | 86% | 0.52 | 377 |
| 1770, Yorkshire 12) | Mary | 22.8% | 84% | 0.34 | 381 | John | 25.6% | 86% | 0.47 | 433 |
| 1825, Yorkshire 8) | Mary | 20.1% | 81% | 0.35 | 99299 | John | 18.8% | 79% | 0.30 | 91111 |

**Table 5 (cont'd)**

| Year, Location | Females | | | | | Males | | | | |
|---|---|---|---|---|---|---|---|---|---|---|
| | Top Name | Pop. | Top 10 Pop. | Top 10 Info I$_s$ | Sample Size | Top Name | Pop. | Top 10 Pop. | Top 10 Info I$_s$ | Sample Size |
| 1350, North/Cumbria 9a) | | | | | | John | 34.5% | 89% | 0.74 | 328 |
| 1530, North/Cumbria 13) | Jane | 16.0% | 84% | 0.24 | 852 | John | 23.1% | 74% | 0.47 | 870 |
| 1550, North/Cumbria 13) | Margaret | 15.6% | 86% | 0.23 | 1491 | John | 21.7% | 75% | 0.40 | 1515 |
| 1580, North/Cumbria 13) | Margaret | 16.8% | 84% | 0.24 | 3750 | John | 18.0% | 71% | 0.32 | 3765 |
| 1610, North/Cumbria 13) | Elizabet | 15.8% | 84% | 0.26 | 4000 | John | 18.2% | 74% | 0.32 | 4044 |
| 1640, North/Cumbria 13) | Elizabet | 16.6% | 87% | 0.30 | 2888 | John | 19.7% | 75% | 0.40 | 2914 |
| 1670, North/Cumbria 13) | Elizabet | 16.5% | 86% | 0.31 | 3813 | John | 19.6% | 75% | 0.40 | 3834 |
| 1700, North/Cumbria 13) | Ann | 16.4% | 86% | 0.35 | 3064 | John | 21.1% | 77% | 0.44 | 3070 |
| 1730, North/Cumbria 13) | Ann | 18.1% | 87% | 0.36 | 2038 | John | 21.6% | 80% | 0.41 | 2038 |
| 1760, North/Cumbria 13) | Ann | 18.8% | 89% | 0.35 | 2830 | John | 23.2% | 81% | 0.44 | 2830 |
| 1790, North/Cumbria 13) | Mary | 19.4% | 89% | 0.37 | 2139 | John | 23.4% | 83% | 0.48 | 2141 |
| 1825, North/Cumbria 8) | Mary | 20.3% | 88% | 0.29 | 24857 | John | 21.8% | 85% | 0.37 | 21966 |
| 1350, Hereford 9b) | Alice | 21.9% | 84% | 0.34 | 576 | John | 34.8% | 89% | 0.57 | 2066 |
| 1700, Hereford 14) | | | | | | John | 20.3% | 78% | 0.44 | 931 |
| 1825, Hereford 8) | Mary | 21.7% | 85% | 0.47 | 6832 | John | 18.9% | 90% | 0.31 | 6350 |
| 1280, East Anglia 15) | | | | | | John | 22.3% | 74% | 0.45 | 391 |
| 1400, East Anglia 15) | | | | | | John | 36.1% | 90% | 0.80 | 590 |
| 1385, soldiers 19) | | | | | | John | 28.1% | 84% | 0.60 | 829 |
| 1550, sailors 16) | | | | | | John | 21.4% | 70% | 0.35 | 583 |
| 1560, Canterbury 17) | Elizabet | 13.6% | 74% | 0.18 | 661 | John | 20.3% | 75% | 0.42 | 5986 |
| 1560, Gloucester 18) | Joan | 18.7% | 88% | 0.25 | c. 4000 | John | 21.4% | 80% | 0.53 | c. 4000 |

Note: The data refer to persons named (born) in a generation about the year given. For details and sources, see Appendix D.

About the year 1300, a high point of medieval economic development, the popularity of the most popular names was similar to that about the year 1800.[72] Since the popularity of the most popular names about 1300 is similar to that early in the sixteenth century and the Black Death decreased name personalization, there must have been some increase in personalization across the fifteenth century.[73] The extent of name personalization shows some fluctuations from early in the sixteenth century to the end of the eighteenth century, but there is no overall trend. Name personalization for males decreased from the early sixteenth to early seventeenth centuries, while little change occurred for females.[74] Name personalization subsequently rose for both males and females. One period of relatively rapid change appears to have been the early to mid seventeenth century.[75]

While more research might be able to explicate the social and economic forces shaping these changes, a broad contrast is clear. Long-term secular changes in agricultural productivity, urbanization, and commercialization before 1800 laid the foundation for the industrial revolution in England. In contrast, the information economy, at least with respect to naming, changed in a less directed way before the nineteenth century. Figures of 20%, 80%, and 0.4 are representative for top name popularity, top ten name popularity, and social information $I_s$, respectively, from 1500 to 1800, as well as circa 1300. As the changes in the most popular female name suggest, this constancy in the extent of name personalization did not reflect stable preferences over a fixed set of names (Mary did not become the most popular female name until the eighteenth century). The trend toward name personalization over the past two hundred years is a significant change relative to the trend over the previous five hundred years. Setting aside the name dynamics that the Norman Conquest created, the reduction in popularity of the most popular names over the past two hundred years appears distinctive relative to the previous thousand years of naming history.

### D. Further Insights from Disciplined Description

Important changes in communications technology in England before 1800 did not generate lasting changes in naming. The spread of printing presses has been described as a key agent of a "massive and decisive cultural 'change of phase' that occurred five centuries ago."[76] The growth of broadsides from early in the sixteenth century and the growth of large public meetings from the eighteenth century were also important communications developments in England.[77] Yet statistics on name popularities suggest that these developments were not sufficient to change the amount of information in the over-all distribution of names, i.e. the extent of shared symbolic experience in naming.

---

[72] Compare 1300, Lincoln and 1290, London in Table 5 to the data for 1800 in Table 3.

[73] In Table 5, compare 1300, Lincoln and 1290, London to 1510, London and 1530, North/Cumbria.

[74] Compare 1510, London to 1610, London, and 1530, North/Cumbria to 1610, North/Cumbria.

[75] Compare 1610, North/Cumbria to 1640, North/Cumbrian, and 1620, Yorkshire to 1670, Yorkshire.

[76] Eisenstein (1980) Vol. II, p 702, 704.

[77] For discussion of these developments, see Hindley (1871/1969), Foreword, and Jephson (1892/1968).

In both the UK and the US, the growth of mass media also did not drive changes in the extent of shared symbolic experiences in naming. Dramatic increases in name personalization were occurring by the mid-nineteenth century, before radio, and television, and large newspaper companies. In contrast, the popularity of the top name, the popularity of the top ten names, and social information $I_s$ all changed little in the first half of the twentieth century, a period in which the newspaper and magazine business grew significantly in scale and scope. Overall, the end of the twentieth century features the production and distribution of common packages of symbols on a scale scarcely imaginable as little as a century ago. It also features name personalization to an extent unprecedented in at least a millennium. There does not, however, appear to be a strong connection between these two contemporary features of the information economy. At least with respect to names, mass media appear to be relatively unimportant in shaping the extent of shared symbolic experience.

The similarity of developments in naming in the US and UK over the past two hundred years should be appreciated in light of important differences between the countries. First, consider geography. The UK consists mainly of a small island with many settlements that have a long and relatively continuous cultural history. The US spans a large continent in which long-distance immigrants established many new settlements. Second, consider population. The population of the US was about 70% of the population of England and Wales in the early nineteenth century, but it grew to five times the population of England and Wales by the end of the twentieth century. Third, consider economic growth. The UK was the first industrial country, and it had acquired a global empire by the end of the nineteenth century. At that point, the US, relative to the UK, was a less developed economy. In the twentieth century the US grew much faster than the UK, and the US become the world's leading economy.

Given these differences, the similarity of personalization trends in the US and the UK is significant. Personalization appears to be an aspect of personal preferences relatively invariant to population size, geography, and income. Further research might explore what caused the trend of increasing personalization. This paper merely provides a historical description[78]: a new trend toward personalization appears as a major change about the early nineteenth century, at least among persons sharing the English language and much European culture.

## III. Trends in Effective Communication

For the past three decades or longer, industrial-scale symbolic production and distribution has been thought to be reshaping information economies, along with all of society. In 1969 the dean of a leading US school of communication declared:

> *In only two decades of massive national existence television has transformed the*
> *political life of the nation, has changed the daily habits of our people, has molded*
> *the style of the generation, made overnight global phenomena out of local*

---

[78] As Mandelbrot (1983) pp. 422-4 points out, good descriptions can be more important than explanatory models for understanding truth and consequences.

> *happenings, redirected the flow of information and values from traditional channels into centralized networks reaching into every home. In other words, it has profoundly affected what we call the process of socialization, the process by which members of our species become human.*[79]

Names such as the "age of information" (1971), "post-industrial society" (1973), "information revolution" (1974) and "communications age" (1975) show recognition of the importance of symbol production.[80] The novelty of these phenomena, however, remains contested through the contest over naming. As one analyst noted in the mid 1990s:

> *Contemporary culture is manifestly more heavily information laden than any of its predecessors. We exist in a media saturated environment….Experientially this idea of an "information society' is easily enough recognised, but as a definition of a new society it is considerably more wayward than any of the notions we have considered.*[81]

From this perspective, a lack of attention to the moral dimension or "master idea" of daily life deprives information of meaning and purpose.[82] This lacuna "must make one deeply sceptical of the 'information society' scenario (while not for a moment doubting that there has been an extensive 'informatisation' of life)…."[83]

Suppose, however, that one suffers from doubts that direct personal experience does not assuage, and one seeks systematic quantitative evidence concerning the extent to which communication has changed. Unfortunately there is no standard approach to quantifying

---

[79] Testimony of George Gerbner, Dean of the Annenberg School of Communication, before the National Commission on the Causes and Prevention of Violence, October 1969, quoted in Hiebert, Ungurait, and Bohn (1974) p. 51-2.

[80] In addition to those given, Beniger (1986) p. 4 lists 71 other related modern societal transformations that scholars have identified since 1950. Beniger himself provides a detailed historical account of crisis, revolution, resolution of crisis, continuity and dynamics stretching back to the Industrial Revolution. His account emphasizes control (that word is used in the title of each of the 10 chapters of the book). The following summarizes the structure of the narrative:

> In only recent years have the industrial economies of the United States and perhaps a dozen other advanced industrial nations appeared to give way to information societies. If this great societal transformation owes its origin to the Industrial Revolution and resulting crisis of control, as argued in the last chapter, why has the resolution of the crisis – the Control Revolution in information-processing and communication technology – continued unabated to this day, almost a century later?
>
> The smooth transition from control crisis to Control Revolution in the 1880s and 1890s can be attributed to three primary dynamics, each of which has sustained the steady development of information societies through the twentieth century.

See Beniger (1986) p. 291.

[81] Webster (1995) p. 22, 23.

[82] Much recent scholarly work has focused on the social construction of meaning, the social regulation of meaning, or the regulation of social meaning. For the most part the literature has not advanced beyond ideas and issues discussed in the early 20'th century advertising industry literature. See Lessig (1995). Analysts of the social construction of meaning, like the stereotype of construction workers, tend to be pre-occupied with pornography and homosexuality.

[83] Ibid, p. 29. Luhmann (1998) p. 89 states, "It is a well-known fact that communication has increased in volume, complexity, memory, and pace." What's behind all this? While the introductory phrase, "It is a well-known fact" might appear otiose, it is not merely stylistic. The phrase is a characteristic instrument of domination that has exerted tremendous symbolic power in this field.

communication from an aggregate, socio-economic perspective. About forty years ago a scholar put forward an extensive theory for analyzing urban growth in terms of information flows. He defined the term "hubits," a bit of information received by a single human being, to enable aggregation of information flows across persons.[84] Information flows were predominately associated with media, and reading was estimated to communicate more than five times as much information as talking and observing the environment. A study fifteen years ago constructed for Japan and the US a census of communication and attempted to measure the supply and consumption of media in word equivalents.[85] Total words consumed were estimated to be only a few percent of total words supplied. A recent study estimated US and world production of information in bytes, and estimated as well as the total amount of information stored in media such as paper, film, and optical and magnetic disks.[86] This study did not explore the significance of the measured information or how it relates to actions.

For insights into communication that has been very socio-economically significant, consider how Islam, Judaism, and Christianity express God's communication with humanity. In Islam, God's message refers to itself as the Qur'an and the Book, the former term associated with recited words and the latter with written words. Communication is a call for response and change: "O you, who believe, respond to Allah and His messenger as they call you to that which gives you life."[87] The response concerns not merely the mind and consciousness, but also a physical change:

> Allah has revealed the most beautiful message in the form of a book...The skins of those who fear their Lord tremble because of it; then their skins and their hearts do soften to the remembrance of Allah...[88]

Hebrew scriptures also emphasize that the word of God effects intentional change in the world. According to Isaiah, the Lord declares:

> As the rain and the snow come down from heaven, and do not return to it without watering the earth and making it bud and flourish, so that it yields seed for the sower and bread for the eater, so is my word that goes out from my mouth: It will not return to me empty, but will accomplish what I desire and achieve the purpose for which I sent it.[89]

For Christians, sacred communication has been actualized in a person:

> In the beginning was the Word, and the Word was with God, and the Word was God....The Word became flesh and made his dwelling among us. We have seen his glory, the glory of the One and Only...[90]

Islam, Judaism, and Christianity are all known as religions of "people of the book," the book that is revered as a key source of knowledge about the same, single God. In these religions, however, communication means much more than reading. Communication is broadband multimedia of the most fantastic sort, and God's word is understood to be intimately related to changes in persons and the world.[91]

---

[84] Meier (1962) p. 131.
[85] Pool (1984).
[86] Lyman and Varian (2000).
[87] Surat 8, al-Anfaal 24.
[88] Surat 39, al-Zumar 23.
[89] Isaiah 55:10-11.
[90] John 1:1, 1:14.

Changes in the popularity of specific personal names indicate effective communication in this broad sense: changes in symbols personally enacted, incarnated, and embedded in ongoing personal interactions. Physical travel of persons, interpersonal oral networks, performers and performances, and media such as books, newspapers, and television all potentially influence the diffusion of names and knowledge about their values. Changes in occupational structures, urban structures, religious beliefs, and family patterns also affect naming norms, and hence can be consider part of communication in a broad sense. Quantifying communication as messages or media growth points to further difficult research into the effects of messages or media. Quantifying communication as changes in important symbolic choices provides a more direct description of the information economy.

This quantification of communication also identifies communication as change in relation to a specific, historical order defining a related type of information. The previous section explored the patterns of name popularity that exist independent of the specific names that constitute the pattern. These patterns were interpreted as expressing social information associated with naming, and changes in social information in this sense have occurred over time. Changes have also occurred in the particular names that configure these patterns. This section explores those changes.[92]

## A. Measuring Effective Name Communication

One measure of effective name communication is the number of new names that appear on successive lists of most popular names. For example, consider lists of the ten most popular names in a population in 1950 and 1960. One could count how many names appear on the 1960 list but not on the 1950 list. An important advantage of this statistic is that it can be calculated using only short name popularity rankings. Studies of names thus need only communicate a small amount of information in order to make such calculations feasible for scholars in different times and places.

The number of new names can be divided by the number of years separating the compared lists to compute name turnover per annum. A recent study uses this statistic. It computes turnover per annum based on name lists ranging from the top 10 to the top 50 most popular names, and on intervals between lists that appear to range from 2 years to

---

[91] For an insightful analysis beyond contemporary bounds, see Ong (1967) Chapter 6. Persons unwilling to take seriously the significance of Islam, Judaism, and Christianity might consider Shakespeare. Shakespeare's sonnets emphasize the power of words, in this case in verse form, to create life and transform temporal personal characteristics into eternal ones (see in particular sonnets 18, 19, 60, 63, 65, and 81; available on the web at http://www.island-of-freedom.com/SHSPEARE.HTM).
[92] The overall pattern of name popularity is related to the stochastic process that governs the movement of individual names. The parameters of this process, however, are complex and difficult to interpret. Separating analysis of the pattern from analysis of changes in its elements should be understood as an interpretive pragmatic, not a theoretical assertion.

100 years.[93]  Turnover per annum is a better statistic than the number of new names if turnover per annum reflects some interpretable model of continual change that encompasses the number of new names measured over different time intervals.  This does not appear to be the case.

The number of new names has major weaknesses as a statistic, and turnover per annum exacerbates these weaknesses.  The number of new names is a rank-based measure that does not consider the magnitude of changes in name popularities.   Limiting in this way the information considered makes sense only given a clear understanding of error processes that justify such a limitation.  Moreover, the number of new names is not independent of the size of the list, the time between lists, or, given any additive error in popularities, the popularity of names on the list.  In addition, the probability of an additional new name appearing on a list is directly related to the number of new names that already appear on the list.  The significance of all these issues is obscured in the turnover per annum statistic, which does not fix list size and averages the number of new names across a varying number of years.[94]

Information theory points to better statistics for measuring changes in name popularity.  Consider lists of name popularities in year 1 and in subsequent year 2.  Let $p_j^i$ represent the popularities of the most popular names in year 2 for lists from years i=1,2 and name ranks j=1,…10.  Let $T^{2,1}$ and $T^{2,2}$ be the total popularity in year 1 and year 2, respectively, of the ten most popular names in year 2:

$$(2) \qquad T^{2,i} = \sum_{j=1}^{10} p_j^i$$

Define communication statistic $C_1$ as the amount of information associated with the change to the popularity of the ten most popular names and "other names":

$$(3) \qquad C_1(p^1, p^2) = \sum_{j=1}^{10} p_j^2 \log_2 \frac{p_j^2}{p_j^1} + (1 - T^{2,2}) \log_2 \frac{1 - T^{2,2}}{1 - T^{2,1}}$$

$C_1$, which has units of bits, is always non-negative and is zero if and only if  $p^1 = p^2$ .

As a communication statistic, $C_1$ has some general limitations.  It measures differences between name probabilities at two points in time, not the flow of information that changes name probability.  Thus if from year 1 to year 2 the name Mary gains popularity and then loses exactly the same amount of popularity, $C_1$ will measure no communication.  Moreover, in some cases $C_1(p^1, p^2) \neq C_1(p^2, p^1)$ .  For a third year with name probabilities  $p^3$ , it may also be the case that $C_1(p^1, p^3) > C_1(p^1, p^2) + C_1(p^2, p^3)$ .  These properties suggest that it is not useful to

---

compute an annual rate of change based on $C_1$. Instead, $C_1$ should be calculated with respect to clearly specified years and interpreted in conjunction with other evidence regarding changes in name popularities.

$C_1$ also has some limitations related specifically to name popularity. As Table 1 shows, the total popularity of the ten most popular names has changed significantly over the past two hundred years. $C_1$ measured across a decade in the beginning of the nineteenth century will differ from $C_1$ measured across a decade at the end of the twentieth century partly because much different weights are attached to the "other names" category.

To avoid the effects of such changes in the "other names" category, probabilities can be normalized across the top ten names. Define communications statistic $C_2$:

$$(4) \quad C_2 = \sum_{j=1}^{10} \frac{p_j^2}{T^{2,2}} \log_2 \frac{p_j^2}{p_j^1} + \log_2 \frac{T^{2,1}}{T^{2,2}}$$

$C_2$ has properties like that of $C_1$, but $C_2$ measures the amount of information associated with changes to the relative popularities of the ten most popular names in year 2. $C_2$ does not depend on the popularity of the "other names" category.

While $C_1$ and $C_2$ are forms of a well-recognized information-theoretic statistic,[95] one of a slightly different type offers an insightful alternative. Note that

$$(5) \quad T^{1,1} \geq T^{2,1} = \sum_{j=1}^{10} p_j^1$$

because the ten most popular names in year 1 may be different from the ten most popular names in year 2. Define communication statistic $C_3$ by replacing $T^{2,1}$ with $T^{1,1}$ in equation (4) above. Thus

$$(6) \quad C_3 = \sum_{j=1}^{10} \frac{p_j^2}{T^{2,2}} \log_2 \frac{p_j^2}{p_j^1} + \log_2 \frac{T^{1,1}}{T^{2,2}}$$

$C_3$ has some appealing properties. From equations (4) and (5), $C_3 \geq C_2$, and if the top ten names are the same in year 1 and year 2, $C_3 = C_2$. $C_3$ rises above $C_2$ as the popularity share of new names in the top ten names increases. Thus $C_3$ measures increased popularity associated with new names as increased communication relative to measure $C_2$. Now consider the total popularity of the top ten names. If it doesn't change from year 1 to year 2, $\log \frac{T^{1,1}}{T^{2,2}} = 0$. Then $C_3$, for small changes in name popularities, approximates a weighted average of the proportional changes in the top ten name popularities. $C_3$ indicates more communication relative to that weighted average if probability flows out of the top ten names from year 1 to year 2, and less if the reverse occurs. Thus $C_3$ associates increased name personalization with increased communication. This association is reasonable in light of the aggregate trends described in Section II.

---

[95] In terms of information theory, $C_1$ and $C_2$ are measures of relative entropy known as the Kullback Leibler distance or the Kullback Leibler measure of directed divergence. See Cover and Thomas (1991) pp. 18-19.

To supplement the above information-theoretic statistics, this paper will also consider an average deviation statistic. Define communication statistic $C_4$:

$$(7) \quad C_4 = 100 \sum_{j=1}^{10} \frac{p_j^2}{T^{2,2}} \left| \frac{p_j^2}{p_j^1} - 1 \right|$$

$C_4$ is the weighted average absolute percentage change in name popularity for the ten most popular names in year 2.

While the number of new names and communication measure $C_4$ are useful supplements, this paper will focus primarily on information-theoretic measures of communication. The nature of these statistics, which provide measurements in bits, can be understood as follows. Suppose that a person in year 2 knows which names are the ten most popular names, but she knows the popularity of these names only as of year 1. $C_1$, $C_2$, and $C_3$ measure, with some variations in scope, the amount of information communicated to her by informing her of the popularities of these names in year 2. This relatively general interpretation of $C_1$, $C_2$, and $C_3$, and their meaningful scale, may help them to contribute to more general understanding of the development of the information economy.

To allow useful compilation and analysis of different studies, the communication statistics $C_1$, $C_2$, $C_3$, and $C_4$ require more extensive reporting from name studies than short name popularity rankings. Calculating these statistics requires information on the popularities on the ten most popular names taken from a different sample. Communicating widely large lists of name popularities is not practical using print, but this can be done easily using the Internet. For many of the name samples analyzed here, sufficient name information has been made available on the Internet for additional studies to calculate the statistics $C_1$, $C_2$, $C_3$, and $C_4$ with respect to new name samples.[96] Other scholars working with name statistics should recognize the importance of this reporting practice for promoting the development of knowledge.

## B. The Shape of Change

Over ten year periods covering the past two centuries, effective communication of new names has changed most dramatically roughly from the beginning of World War I to the end of World War II. Tables 6 and 7 present name communication statistics covering the UK and the US over the past two centuries. These statistics show considerable decade-to-decade variation, as well as differences among the different statistics.[97] For both males and females in the US, a general trend increase in effective communication appears to have occurred from 1915 to 1945. The data for the UK is more fragmentary. For UK females and males, name communication across decades increased significantly sometime between 1880 and 1944. Evidence from twenty-five year intervals suggests that the increase happened between 1900 and 1950, with the increase coming later and more sharply for males than for females. These measures suggest that information economies were transformed in the first half of the twentieth century, rather than with the growth of mass media in the second half of the twentieth century.

---

[96] See the online links in References and, more generally, the AGNAMES site, http://users.erols.com/dgalbi/names/agnames.htm

---

[97] For further data and discussion of variability, see Appendix B.

## Table 6: Name Communication Statistics for the UK

| | Females | | | | | Males | | | | | |
|---|---|---|---|---|---|---|---|---|---|---|---|
| Ending Year | $C_1$ | $C_2$ | $C_3$ | $C_4$ | New Top 10 | $C_1$ | $C_2$ | $C_3$ | $C_4$ | New Top 10 | Ending Year |
| | | | | | *comparisons across 10-year intervals* | | | | | | |
| 1810 | 0.0063 | 0.0044 | 0.0065 | 6% | 1 | 0.0110 | 0.0087 | 0.0176 | 10% | 1 | 1810 |
| 1820 | 0.0055 | 0.0039 | 0.0130 | 7% | 1 | 0.0037 | 0.0039 | 0.0039 | 6% | 0 | 1820 |
| 1830 | 0.0111 | 0.0147 | 0.0249 | 9% | 1 | 0.0042 | 0.0026 | 0.0026 | 6% | 0 | 1830 |
| 1840 | 0.0075 | 0.0097 | 0.0097 | 8% | 0 | 0.0021 | 0.0010 | 0.0074 | 4% | 1 | 1840 |
| 1850 | 0.0080 | 0.0065 | 0.0065 | 9% | 0 | 0.0030 | 0.0013 | 0.0013 | 4% | 0 | 1850 |
| 1860 | 0.0156 | 0.0200 | 0.0475 | 13% | 2 | 0.0111 | 0.0103 | 0.0243 | 10% | 1 | 1860 |
| 1870 | 0.0250 | 0.0137 | 0.0137 | 15% | 0 | 0.0196 | 0.0159 | 0.0330 | 14% | 1 | 1870 |
| 1880 | 0.0380 | 0.0558 | 0.1063 | 27% | 2 | 0.0096 | 0.0050 | 0.0050 | 10% | 0 | 1880 |
| | | | | | *comparisons across 25-yer intervals* | | | | | | |
| 1875 | 0.3558 | 0.6486 | 0.8787 | 464% | 4 | 0.0856 | 0.1179 | 0.1655 | 45% | 2 | 1875 |
| 1900 | 0.4002 | 1.0401 | 1.4859 | 4237% | 4 | 0.0342 | 0.0236 | 0.0654 | 20% | 2 | 1900 |
| 1925 | 0.5482 | 1.3097 | 1.7903 | 25099% | 5 | 0.1772 | 0.4475 | 0.6923 | 217% | 3 | 1925 |
| 1950 | 1.0529 | 2.3962 | 3.5065 | 3709% | 7 | 0.6538 | 0.8054 | 2.0464 | 723% | 8 | 1950 |
| 1975 | 0.7826 | 1.2914 | 3.7448 | 2753% | 9 | 0.4838 | 1.2094 | 1.9572 | 989% | 6 | 1975 |
| | | | | | *comparisons across 10-year intervals* | | | | | | |
| 1954 | 0.0769 | 0.2322 | 0.2822 | 69% | 1 | 0.1361 | 0.3573 | 0.5100 | 151% | 3 | 1954 |
| 1964 | 0.1887 | 0.5729 | 1.2198 | 234% | 6 | 0.1115 | 0.2582 | 0.4197 | 103% | 5 | 1964 |
| 1974 | 0.2196 | 0.2984 | 1.5670 | 282% | 9 | 0.0287 | 0.0756 | 0.1853 | 31% | 3 | 1974 |
| 1984 | 0.1769 | 0.5852 | 1.0152 | 291% | 4 | 0.0348 | 0.1042 | 0.2252 | 39% | 3 | 1984 |
| 1994 | 0.0903 | 0.3096 | 0.9310 | 120% | 6 | 0.2014 | 0.5368 | 1.3947 | 326% | 7 | 1994 |
| 1800-1880 ave. | 0.0146 | 0.0161 | 0.0285 | 12% | 0.9 | 0.0080 | 0.0061 | 0.0119 | 8% | 0.5 | |
| 1944-1994 ave. | 0.1505 | 0.3997 | 1.0030 | 199% | 5.2 | 0.1025 | 0.2664 | 0.5470 | 130% | 4.2 | |
| Ratio | 10 | 25 | 35 | 17 | 6 | 13 | 44 | 46 | 16 | 8 | |

## Table 7: Name Communication Statistics for the US
### (on a decade-to-decade basis)

| Ending Decade Midpoint | Females | | | | | Males | | | | | Ending Dec. |
|---|---|---|---|---|---|---|---|---|---|---|---|
| | $C_1$ | $C_2$ | $C_3$ | $C_4$ | New Top 10 | $C_1$ | $C_2$ | $C_3$ | $C_4$ | New Top 10 | |
| 1815 | 0.0031 | 0.0052 | 0.0107 | 6% | 1 | 0.0061 | 0.0088 | 0.0280 | 11% | 1 | 1815 |
| 1825 | 0.0054 | 0.0094 | 0.0094 | 9% | 0 | 0.0037 | 0.0076 | 0.0076 | 6% | 0 | 1825 |
| 1835 | 0.0026 | 0.0028 | 0.0028 | 5% | 0 | 0.0018 | 0.0030 | 0.0030 | 6% | 0 | 1835 |
| 1845 | 0.0064 | 0.0124 | 0.0365 | 10% | 1.5 | 0.0039 | 0.0048 | 0.0174 | 7% | 1 | 1845 |
| 1855 | 0.0164 | 0.0319 | 0.0353 | 18% | 1 | 0.0075 | 0.0140 | 0.0304 | 10% | 1 | 1855 |
| 1865 | 0.0136 | 0.0217 | 0.0278 | 16% | 1 | 0.0050 | 0.0099 | 0.0099 | 11% | 0 | 1865 |
| 1875 | 0.0118 | 0.0113 | 0.0475 | 15% | 1.5 | 0.0074 | 0.0027 | 0.0027 | 10% | 1 | 1875 |
| 1885 | 0.0098 | 0.0114 | 0.0357 | 14% | 1 | 0.0154 | 0.0124 | 0.0361 | 16% | 1 | 1885 |
| 1895 | 0.0058 | 0.0135 | 0.0450 | 12% | 1 | 0.0151 | 0.0084 | 0.0145 | 17% | 1 | 1895 |
| 1905 | 0.0038 | 0.0127 | 0.0345 | 12% | 2 | 0.0089 | 0.0070 | 0.0070 | 14% | 0 | 1905 |
| 1915 | 0.0063 | 0.0171 | 0.0286 | 12% | 1 | 0.0051 | 0.0157 | 0.0157 | 11% | 1 | 1915 |
| 1925 | 0.0326 | 0.1072 | 0.1940 | 42% | 2 | 0.0161 | 0.0374 | 0.0752 | 23% | 1 | 1925 |
| 1935 | 0.0818 | 0.2482 | 0.5404 | 93% | 4 | 0.0135 | 0.0277 | 0.0416 | 20% | 2 | 1935 |
| 1945 | 0.1505 | 0.4070 | 0.8016 | 213% | 5 | 0.0354 | 0.0902 | 0.1929 | 38% | 2 | 1945 |
| 1955 | 0.1342 | 0.4662 | 0.6515 | 208% | 2 | 0.0279 | 0.0787 | 0.1267 | 34% | 1 | 1955 |
| 1965 | 0.0574 | 0.2513 | 0.4955 | 72% | 5 | 0.0111 | 0.0320 | 0.0885 | 17% | 2 | 1965 |
| 1975 | 0.0719 | 0.3149 | 0.6161 | 107% | 5 | 0.0803 | 0.2886 | 0.4712 | 134% | 4 | 1975 |
| 1985 | 0.0909 | 0.3945 | 0.7494 | 205% | 5 | 0.0182 | 0.0737 | 0.2297 | 29% | 3 | 1985 |
| 1995 | 0.0195 | 0.1200 | 0.4298 | 43% | 4 | 0.0193 | 0.0626 | 0.1888 | 32% | 2 | 1995 |
| 1805-1915 ave. | 0.0077 | 0.0136 | 0.0285 | 12% | 1.0 | 0.0073 | 0.0086 | 0.0157 | 11% | 0.5 | |
| 1945-1995 ave. | 0.0748 | 0.3094 | 0.5885 | 127% | 4.2 | 0.0313 | 0.1067 | 0.2210 | 49% | 2.4 | |
| Ratio | 10 | 23 | 21 | 11 | 4 | 4 | 12 | 14 | 5 | 5 | |

The amount of effective communication has risen significantly relative to the amount of information in the naming distribution. As noted above, $C_2$ and $C_3$ provide measures in bits, and these statistics are not affected by the secular fall in the popularity of the ten most popular names. Hence these statistics probably provide the most useful long-term comparisons. They show that effective communication increased by about factors of 22 and 13 for US females and males, respectively, and by factors of 30 and 45 for UK females and males, respectively.[98] In contrast, the amount of social information in the naming distribution decreased by factors of 10 and 15 for US females and males, respectively, and factors of 5 and 10 for UK females and males, respectively.[99] In absolute terms, in the beginning of the nineteenth century the amount of information in the naming distribution was large relative to changes in names on a decade-to-decade basis. At the end of the twentieth century the opposite is true.

The magnitude of effective name communication differs significantly between the US and the UK, and between females and males. At the end of the twentieth century effective name communication in the US averages about one-half as much as that in the UK.[100] This difference might be attributable to factors traditionally considered significant to communication: the US has a larger, more geographically dispersed and more culturally heterogeneous population than the UK. Effective communication of new names is also about two to three times higher for females than for males. This might reflect more familial inertia in men's names, perhaps resulting from more importance being attached to naming males after older male kin. The difference might also indicate more extensive social networks among females and their value in effective communication.

Effective communication differs significantly over different time horizons. Table 8 shows name communication statistics over periods of a hundred years or longer. In the US name communication in the twentieth century was roughly twice that in the nineteenth. For the UK, female name communication seems to have fallen from the nineteenth to the twentieth century, while male name communication shows no clear trend. These figures contrast sharply with increases by factors of 13 to 45 for name communication on a decade-to-decade basis. Effective name communication does not seem to aggregate simply; more change in the short term does not necessarily imply more long-term change.[101] From a long-term perspective, slow and steady change can be equivalent to rapid short-term change that dies out or cycles over time.[102]

---

[98] These figures are based on the ratios of the decade-to-decade averages given on the bottom of Tables 6 and 7. The differences between $C_2$ and $C_3$ are relative small. The figures in the text above are averages for the ratios for these two statistics.

[99] These figures are based on the ratio of estimated trend values for $I_s$ about 1805 and 1995. See Tables 3 and 4.

[100] Based on comparison of late-twentieth century averages given at the bottom of Tables 6 and 7.

---

[101] About 1875 changes in $C_2$, $C_3$, $C_4$, and new names on a decade-to-decade basis are less than twice as large as changes on a year-to-year basis. See Appendix B.

[102] These results thus appear to support Daniel Bell's position with respect to cultural change: "Technologies have more to do with the speed of change than with the character of the changes." Bell (1973/1999) forward, lii ft. 24.

## Table 8
## Name Communication Over a Century or Longer

| Time Span, Place | Females | | | | | Males | | | | |
|---|---|---|---|---|---|---|---|---|---|---|
| | $C_1$ | $C_2$ | $C_3$ | $C_4$ | New Top 10 | $C_1$ | $C_2$ | $C_3$ | $C_4$ | New Top 10 |
| 1066-1166, Conquest Model | | | | | | 1.083 | | 4.117 | 982% | 10 |
| 1080-1180 Winchester | | | | | | 0.508 | 0.377 | 2.303 | 367% | 5 |
| 1290-1510, London | | | | | | 0.446 | 0.510 | 0.804 | 658% | 5 |
| 1510-1610, London | | | | | | 0.294 | 0.406 | 0.518 | 236% | 3 |
| 1610-1825, London | | | | | | 0.312 | 0.292 | 0.433 | 112% | 3 |
| 1350-1610, Manchester | | | | | | 1.058 | 1.379 | 1.647 | 2025% | 4 |
| 1610-1805, Manchester | | | | | | 0.235 | 0.263 | 0.335 | 92% | 2 |
| 1350-1620, Yorkshire | 4.120 | 3.943 | 4.959 | 12660% | 6 | 1.413 | 1.640 | 1.799 | 3819% | 5 |
| 1620-1720, Yorkshire | 0.353 | 0.374 | 0.507 | 320% | 3 | 0.107 | 0.084 | 0.233 | 36% | 2 |
| 1720-1825, Yorkshire | 0.364 | 0.445 | 0.504 | 192% | 3 | 0.415 | 0.525 | 0.665 | 202% | 3 |
| 1350-1525, North/Cumbria | 0.216 | 0.200 | 0.330 | 72% | 1 | 0.341 | 0.438 | 0.586 | 255% | 5 |
| 1525-1625, North/Cumbria | 0.260 | 0.271 | 0.373 | 347% | 2 | 0.030 | 0.035 | 0.047 | 21% | 3 |
| 1625-1725, North/Cumbria | 0.068 | 0.075 | 0.088 | 31% | 1 | 0.058 | 0.055 | 0.099 | 26% | 2 |
| 1725-1825, North/Cumbria | | | | | | 0.081 | 0.071 | 0.098 | 28% | 1 |
| 1350-1700, Hereford | | | | | | 0.939 | 1.180 | 1.500 | 3182% | 4 |
| 1700-1825, Hereford | | | | | | 0.189 | 0.123 | 0.145 | 52% | 3 |

**Table 8 (continued)**

| Time Span, Place | Females | | | | | Males | | | | |
| | $C_1$ | $C_2$ | $C_3$ | $C_4$ | New Top 10 | $C_1$ | $C_2$ | $C_3$ | $C_4$ | New Top 10 |
|---|---|---|---|---|---|---|---|---|---|---|
| 1280-1400, East Anglia | | | | | | 0.198 | 0.038 | 0.110 | 38% | 3 |
| 1800-1900, UK | 1.577 | 3.915 | 4.512 | 21603% | 7 | 1.280 | 2.324 | 2.629 | 12548% | 5 |
| 1850-1954, UK | 0.976 | 2.965 | 3.876 | 135285% | 6 | 0.838 | 1.899 | 3.781 | 1808% | 8 |
| 1900-1994, UK | 0.463 | 1.132 | 3.556 | 5587% | 10 | 0.837 | 2.295 | 4.612 | 11068% | 8 |
| 1805-1905, US | 0.448 | 1.513 | 2.062 | 1753% | 5 | 0.223 | 0.693 | 1.031 | 297% | 5 |
| 1855-1955, US | 0.752 | 2.735 | 3.845 | 8952% | 8 | 0.268 | 0.755 | 1.224 | 207% | 5 |
| 1895-1995, US | 0.467 | 3.276 | 4.054 | 43436% | 8 | 0.320 | 1.674 | 3.065 | 1386% | 7 |

Across a century simple forms of communication can be astonishingly powerful. When the Normans conquered England in 1066, the inhabitants of England rapidly adopted Norman names. Table 8 shows that effective communication of new names following the Norman Conquest was about the same magnitude as that over the twentieth century in England and Wales. The Conquest probably was a powerful impetus to listen, observe, discuss, and travel, particularly given that the conquerors were well-received.[103] The figures for London suggest that subsequent rates of name communications in the Middle Ages and early modern period were significantly lower than in the nineteenth and twentieth centuries (for additional data, see Appendix A, Table A1). The Norman Conquest appears to have spurred an exceptional amount of name communication for its time. Yet there is nothing exceptional about the normal, personal means of communication that brought about the dramatic symbolic change.

## C. Talking about Communication

Effective communication concerns not only media and messages but also personal activity and personal relationships.[104] The most dramatic increase in effective communication of names on a decade-to-decade basis occurred between 1915 and 1945, when mobilization for world wars stirred persons into many new types of activities and reshaped a wide range of organizations.[105] Effective communication of female names has differed significantly from effective communication of male names, a fact consistent with the importance of sex to personal relationships and lived experiences. Over a century, even dramatic differences in social organization and technology may not matter. Human nature, including the capacities of the human body, is such as to enable actions and create personal relationships that can support in widely differing circumstances similar levels of effective communication.[106]

An important aspect of actions and personal relationships is the scope of communicative peers. By the beginning of the sixteenth century, dominant political, commercial, and religious elites had established among themselves a communications system such that, in

---

[103] On the absence of cultural hostility on the part of the English toward the Normans, see Clark (1995b) p. 291. Bell (1973/1999) forward, lii ft. 24 argues: "The greatest forces for the 'flattening' of cultures have been conquest – military, political and religious." The Norman Conquest made names in England much more similar to names in continental Europe. But that the Norman Conquest made the most popular names in England relatively less popular is doubtful. See Table 5.

[104] Brown and Duguid (2000) make this point clearly with many interesting examples.

[105] Careful analysis and classification of jobs shows that the share of information workers grew rapidly in the US in the 1920s. By 1930 information workers were the largest sector in a four-sector (information, agriculture, industry, and services) analysis of the US workforce. See Schement (1990).

[106] Those who are troubled by doubts about the existence of human nature, truth, and reality might benefit from reciting words of Donald Davidson, an eminent contemporary philosopher:

> …I believe in the ordinary notion of truth: there really are people, mountains, camels and stars out there, just as we think there are, and those objects and events frequently have the characteristics we think we perceive them to have. Our concepts are ours, but that doesn't mean they don't truly, as well as useful, describe an objective reality.

See Davidson (1999) p. 19.

two weeks, a message could cross the length and breadth of Europe.[107]   But in the Middle Ages communication also occurred in less contrived ways among a much wider set of communicative peers:

> *...gentry and higher clergy traveled constantly about, and with considerable trains, bringing glimpses and whispers of new fashions to the peasantry; pilgrims set off, and sometimes returned, while others passed through on their way to or from distant shrines; townspeople saw and heard foreign traders of all sorts, and many of them went on trading voyages of their own; peasants often had carrying-services to perform, ...no doubt returning full of the novelities seen and heard on the way.  Thus, townspeople would have been well up-to-date with fashions, and not even the humblest villagers need have been quite unaware of the great world and of its ways of thinking and behaving.[108]*

This much wider set of peers allowed communication to scale better to cover large areas. In the words of a contemporary observer, "Tongues carry tales to every place/Much faster than a coach could race."[109]

While accounts of the spread of Luther's Ninety-Five Theses emphasize the impact of printing technology, that historical incident also shows the power of peer-to-peer networking.  When Luther issued his theses in 1517, they were said to have been known throughout Europe in a month.[110]   In that amount of time an official communication, under good circumstances, might have traveled from Leipzig to Rome to Paris and on to London, leaving of course many other European towns without the information.[111]   Peer-to-peer networking among a new group of peers played a key role in quickly spreading Luther's theses:

> *The educated élite who could understand Latin and theological debate was no longer composed only of churchmen and professors.  [Luther's theses] were initially read by a small group of learned laymen who were less likely to gather on the church steps than in urban workshops where town and gown met to exchange gossip and news, peer over editors' shoulders, check copy and read proof.  There, also, new schemes for promoting bestsellers were being tried out.[112]*

Without any central direction and using only existing social networks and organizations, Luther's theses were translated from Latin into multiple vernaculars, printed in multiple towns, promoted at newsstands, and sold by peddlers.[113]   If only the Roman Catholic Church ordered the printing and distribution of Luther's theses, Luther's arguments

---

[107] Aston (1968) p. 57.
[108] Clark (1995a) p. 79.
[109] Aston (1969) p. 58.
[110] Ibid p. 76.
[111] Based on evidence and discussion in ibid pp. 54-60.
[112] Eisenstein, vol. I, pp. 308-9.
[113] Ibid.  This process depended on a loose degree of control over the work, and it produced some results that annoyed Luther:

> One need only check the places of the reprints of Luther's tracts between 1518 and 1522 to note the geographic dimension of a printing enterprise that nowadays would be severely handicapped by the laws of copyright.  The printers of Wittenberg at times even published material that Luther did not want to have published.  This aspect of the matter annoyed him no end, but on the other hand he was glad to have their services…

From Hillerbrand (1968) p. 275.

probably would have been no more widely communicated than the decrees and canons of the Council of Trent.[114]

## IV. Conclusions

Scientific study of personal names provides important insights into personalization and effective communication. A strong trend toward name personalization developed at the beginning of the nineteenth century, before the rise of mass media. Moreover, effective communication of names has always been strongly related to actions and personal relationships. The Norman Conquest, sex differences, and world wars have significantly shaped effective communication of names. Increased personal exposure to television and the growth of large media companies in the second half of the twentieth century have not dramatically changed trends in personalization and effective communication.[115]

Other evidence in addition to personal names also points to the importance of personalization and its likely growth in the future. Spending on person-to-person communication is much greater than spending for content not produced for a specific person.[116] E-mail and instant messaging, which primarily involve personalized content produced on a non-commercial basis, are widely considered to be the Internet services that persons value most highly.[117] Brand proliferation is a rudimentary form of personalization associated with fee-simple transactions for goods and low bandwidth feedback from consumption to production. While there is evidence of increasing brand proliferation, the growth of more complex service transactions and the development of Internet technologies point toward much more extensive personalization.[118] In fact, firms are investing significantly in technologies to personalize interactions with their customers, and much of the marketing literature now focuses on ways to build personal relationships with customers.[119]

---

[114] The decrees and canons of the Twenty-Fifth Session of the Council of Trent (1563) forbid anyone, under pain of excommunication, "without our authority to publish, in any form, any commentaries, glosses, annotations, scholia, or any kind of interpretation whatsoever of the decrees." The decrees were to be read aloud in two churches in Rome, affixed to doors and gates in four locations, and eventually published in the Roman press. See document at http://history.hanover.edu/texts/trent/ct25.html

[115] As Galbi (2001b) documents, the growth of media since 1925 also has not dramatically affected the share of advertising in GDP or real advertising spending per person-hour of media use. The amount of discretionary time on average a person spends in primary activities other than media use also shows no trend over the past seventy-five years.

[116] Odlyzko (2001).

[117] For example, AT&T President David Dorman recently noted that many customers mainly use the Internet for e-mail, and he pointed to the need for new services to foster industry development. See http://www.zdnet.com/zdnn/stories/news/0,4586,2806602,00.html

[118] For a discussion of trends in brand proliferation, see David (2000) pp. 61-5. Most discussions of new goods from a national accounting perspective concern how new goods and brand proliferation complicate calculation of traditional price indices and productivity measures. However, as this paper suggests, trends in brand names and brand tokens are important economic developments in themselves, and such naming trends deserve more attention. For discussion of tools for coping with product diversity in e-commerce, see Nadel (2000).

[119] For early, insightful business analysis of these developments, see McKenna (1991) and McKenna (1997).

The trend toward personalization has important implications for communications policy. Given that new, diverse, personalized media are likely to create new personal activities and drain personal attention and spending from mass media, the balance of traditional anxieties over diversity versus national unity is likely to shift toward national unity.[120] Moreover, network and local spot television, the traditional primary channels of political competition, will present a shrinking share of relatively inert media consumers while confronting increasingly acute strains related to access and campaign financing.[121] Since attempts to regulate shared symbolic experiences through mass media are impractical and ineffective, and these facts are becoming more apparent, the net effect is likely to be an increase in fear, uncertainty and doubt about the future of communications.[122] Communications policy needs to confront directly such anxieties and to ensure that the value of peer-to-peer information creation and communication is widely experienced.[123]

The evidence in this paper suggests that economic theory needs to focus more attention on relationships among persons. Information flows depend significantly on personal contacts, even in situations in which advanced information and communications technology is available. The discipline of psychology has helped to enrich economic models of individual decision-making. Social psychology and sociology might similarly provide insights into information flows and the network economics of communicative ties among persons. Study of personal networks in an important sense provides the micro-foundations for study of organizations that are increasingly understood as networks of capabilities, rather than as units engaging in anonymous market transactions of a well-specified and narrow sort.

The growth of personalization also has important implications for statistical agencies and for the measurement and monitoring of macro-economies. Changes in the quality and range of goods and services present major challenges to conventional national accounting. The growth of personalized services, customized pricing, highly malleable information goods, and relationship-oriented transactions greatly magnifies these challenges.[124] To help address these developments, statistical agencies need to develop new statistics different from those traditional associated with production and distribution of goods. Personal names offer the advantage of already being collected through major government statistical programs. Names, however, are not currently being compiled and analyzed systematically to measure and monitor developments in the information economy.  Such analysis potentially could contribute at low cost to understanding better the impact of information technology and the development of a "new economy."

---

[120] For some early evidence of such a shift, see Sunstein (2001). This development should be distinguished from any movements in the long-running, highly polarized, generally unscientific debate about trends in media concentration. See Compaine (1999), Compaine (2001), McChesney (2001), Netanel (2000) pp. 468-70, Pastore (2001), Roppen (1997), and Shirkey (2001), response to question 7.

[121] This issue, and a possible policy approach to addressing it, is explored in Galbi (2001a). For a discussion of this development from the perspective of commerce, see Shirky (n.d.).

[122] For an insightful analysis of the institutional context of fear, uncertainty, and doubt, see Irwin (1998).

[123] Benkler (2001) and Benkler (2000) discuss the benefits of expanding personal agency in information creation and communication. Benkler (1998) p. 194 footnote 37 also provides an attempt to assuage anxiety about such a development.

[124] Quah (2001) conceptualizes these developments as a decrease in distance between production and consumption. Technology for customizing prices (see Wessel (2001)) provides a good example of changes in standard forms of economic information.

# References

## Primary Data Sources

# Appendix A
## Additional Data on Name Communication in England Before 1825

### Table A1

| Time Span, Place | Females | | | | | Males | | | | |
|---|---|---|---|---|---|---|---|---|---|---|
| | $C_1$ | $C_2$ | $C_3$ | $C_4$ | New Top 10 | $C_1$ | $C_2$ | $C_3$ | $C_4$ | New Top 10 |
| 1080-1120, Winchester | | | | | | 0.027 | 0.091 | 0.297 | 36% | 3 |
| 1120-1180, Winchester | | | | | | 0.463 | 0.248 | 0.506 | 317% | 5 |
| 1200,1210; Essex, South | 0.340 | 0.224 | 0.535 | 136% | 4 | 0.036 | 0.041 | 0.060 | 25% | 1 |
| 1210,1270; South, Rutland | 0.258 | 0.075 | 0.280 | 57% | 4 | 0.127 | 0.094 | 0.143 | 47% | 2 |
| 1270,1300; Rutland, Lincoln | 0.139 | 0.184 | 0.324 | 67% | 4 | 0.179 | 0.178 | 0.313 | 52% | 2 |
| 1260-1290, London | | | | | | 0.028 | 0.026 | 0.044 | 20% | 1 |
| 1620-1670, Yorkshire | 0.112 | 0.141 | 0.279 | 53% | 4 | 0.065 | 0.039 | 0.039 | 20% | 0 |
| 1670-1720, Yorkshire | 0.258 | 0.225 | 0.282 | 119% | 3 | 0.102 | 0.050 | 0.110 | 33% | 2 |
| 1720-1770, Yorkshire | 0.096 | 0.110 | 0.126 | 42% | 2 | 0.103 | 0.120 | 0.124 | 32% | 1 |
| 1770-1825, Yorkshire | 0.192 | 0.237 | 0.323 | 103% | 2 | 0.222 | 0.281 | 0.390 | 127% | 2 |
| 1550-1580, North/Cumbria | 0.014 | 0.013 | 0.013 | 9% | 0 | 0.014 | 0.013 | 0.013 | 12% | 0 |
| 1580-1610, North/Cumbria | 0.056 | 0.047 | 0.154 | 26% | 1 | 0.011 | 0.012 | 0.012 | 11% | 0 |
| 1610-1640, North/Cumbria | 0.039 | 0.034 | 0.057 | 20% | 1 | 0.007 | 0.008 | 0.008 | 10% | 0 |
| 1640-1670, North/Cumbria | 0.015 | 0.018 | 0.019 | 12% | 1 | 0.013 | 0.016 | 0.050 | 9% | 2 |
| 1670-1700, North/Cumbria | 0.012 | 0.013 | 0.026 | 12% | 1 | 0.006 | 0.006 | 0.018 | 8% | 2 |
| 1700-1730, North/Cumbria | 0.013 | 0.013 | 0.026 | 10% | 1 | 0.009 | 0.006 | 0.012 | 5% | 2 |
| 1730-1760, North/Cumbria | 0.014 | 0.015 | 0.015 | 9% | 1 | 0.006 | 0.006 | 0.009 | 6% | 1 |
| 1760-1790, North/Cumbria | 0.016 | 0.014 | 0.040 | 12% | 1 | 0.012 | 0.008 | 0.046 | 9% | 2 |
| 1790-1825, North/Cumbria | 0.042 | 0.048 | 0.065 | 22% | 2 | 0.033 | 0.035 | 0.039 | 18% | 1 |
| 1385,1550; Soldiers, Sailors | | | | | | 0.161 | 0.151 | 0.191 | 53% | 3 |
| 1550,1560; Sailors, Cantbry | | | | | | 0.040 | 0.042 | 0.042 | 24% | 0 |
| 1560, 1560; Cantbry, Glcstr | 0.204 | 0.122 | 0.157 | 55% | 2 | 0.065 | 0.053 | 0.112 | 27% | 2 |

## Appendix B
## Evidence on Variations in Name Statistics

Analysis of independent samples from closely related populations helps indicate the nature and magnitude of variations in name statistics. Table B1 concerns names of white females born in the US in the 1870s. The first part of the table presents data from the Census of 1880 (females ages 0-9 years), and the second part presents data from the Census of 1920 (females ages 40-49 years). Considering figures for individual years helps to show measurement variance about the long-term trends discussed in Section II A (text and Table 4) and Section III B (text and Table 7). Comparing averages of individual years to figures based on a decade grouping indicates effects of aggregation across years. Comparing samples from the different censuses indicates the importance of changes in recording over historical time and changes in recording in relation to a person's age.[125]

Variations in the name personalization statistics are modest relative to secular trends. Based on a binomial model, sampling variation (standard deviation) for tokens with 10% and 40% probability in a sample size of 4000 are 0.47% and 0.77% of the sample size, respectively. Year-to-year variations in the top name and top ten name probabilities are roughly this size. Female names for females ages 40-49 recorded in the Census of 1920 appear to more personalized, with a lower probability for the top name and less social information in naming. Nonetheless, the long-term trend toward more personalization is discernable in the year-to-year figures across the decade, and the differences between samples are generally less than the effects of the trend over the decade.

The name communication statistics show much more variability. $C_1$, $C_2$, and $C_4$ vary year-to-year by up to a factor of two, and $C_3$ varies up to a factor of five. For $C_2$, $C_3$, $C_4$, and new names, the decade-to-decade figures are no more than twice as large as the average year-to-year figures. These facts suggest that, as least for the US in the 1870s, year-to-year variations in name communication statistics, whether created by coding errors or different aggregate choices, dominate ten-year trends. Across the nineteenth century, however, the communication statistics show much greater change than the year-to-year variability in the 1870s. As these figures indicate, interpreting name communication statistics requires description of the time horizon and discrimination between long-term trends and measurement variability.

More research is needed to describe name communication in a way that more simply and consistently encompasses different time horizons and closely related samples. One would like to better understand what is driving year-to-year variations, and, to the extent that such variation does not reflect interesting aspects of communication, one would like to know how to control for it. One would also like statistics that can scale change over a measured time period to change over other time periods given the same communication process.

---

[125] Blacks faced widespread discrimination in the South following the Civil War. Such discrimination, which may have differed significantly from 1880 to 1920, may have affected the way and extent to which blacks were recorded in the censuses. While this potentially is an interesting and feasible direction of study, here the objective is just to provide some indication of baseline variability, and the sample has been limited to white females.

## Table B1: Variations in Name Statistics

| | Personalization Statistics | | | | Communication Statistics | | | | | | |
| Birth Year | Top Name | Pop. | Top 10 Pop. | Top 10 Info $I_s$ | Year Span | $C_1$ | $C_2$ | $C_3$ | $C_4$ | New Top 10 | Sample Size |
|---|---|---|---|---|---|---|---|---|---|---|---|
| | | | | | Samples from Census of 1880, Ages 0-9 | | | | | | |
| 1871 | Mary | 11.7% | 41.0% | 0.226 | 1870-1871 | 0.0016 | 0.0037 | 0.0037 | 5% | 0 | 4637 |
| 1872 | Mary | 11.6% | 40.2% | 0.239 | 1871-1872 | 0.0024 | 0.0060 | 0.0235 | 7% | 1 | 5043 |
| 1873 | Mary | 10.9% | 39.1% | 0.219 | 1872-1873 | 0.0014 | 0.0030 | 0.0095 | 7% | 1 | 5170 |
| 1874 | Mary | 10.5% | 38.8% | 0.207 | 1873-1874 | 0.0016 | 0.0040 | 0.0040 | 6% | 0 | 5415 |
| 1875 | Mary | 10.7% | 39.4% | 0.202 | 1874-1875 | 0.0012 | 0.0026 | 0.0081 | 4% | 1 | 5339 |
| 1876 | Mary | 10.2% | 37.3% | 0.217 | 1875-1876 | 0.0030 | 0.0041 | 0.0041 | 7% | 0 | 5510 |
| 1877 | Mary | 10.4% | 36.7% | 0.231 | 1876-1877 | 0.0019 | 0.0048 | 0.0048 | 7% | 0 | 5568 |
| 1878 | Mary | 10.6% | 37.5% | 0.232 | 1877-1878 | 0.0020 | 0.0047 | 0.0068 | 7% | 1 | 5748 |
| 1879 | Mary | 9.7% | 35.4% | 0.206 | 1878-1879 | 0.0031 | 0.0057 | 0.0158 | 8% | 1 | 5099 |
| 1880 | Mary | 9.5% | 34.8% | 0.224 | 1879-1880 | 0.0012 | 0.0032 | 0.0048 | 6% | 1 | 5737 |
| ave. years 1870s | Mary | 10.6% | 38.0% | 0.220 | ave. y-to-y 1870s | 0.0019 | 0.0042 | 0.0085 | 6% | 0.6 | 5327 |
| grp 1870s | Mary | 10.2% | 37.3% | 0.206 | grp 1860s to 70s | 0.0130 | 0.0054 | 0.0082 | 14% | 1 | 53258 |
| | | | | | Samples from Census of 1920, Ages 40-49 | | | | | | |
| 1871 | Mary | 10.9% | 41.6% | 0.211 | 1870-1871 | 0.0030 | 0.0068 | 0.0221 | 8% | 2 | 3304 |
| 1872 | Mary | 10.1% | 40.9% | 0.174 | 1871-1872 | 0.0036 | 0.0085 | 0.0085 | 9% | 1 | 3673 |
| 1873 | Mary | 10.4% | 40.2% | 0.198 | 1872-1873 | 0.0041 | 0.0102 | 0.0266 | 9% | 1 | 3401 |
| 1874 | Mary | 9.8% | 39.2% | 0.180 | 1873-1874 | 0.0034 | 0.0086 | 0.0493 | 9% | 2 | 3496 |
| 1875 | Mary | 10.2% | 38.4% | 0.191 | 1874-1875 | 0.0035 | 0.0087 | 0.0118 | 9% | 1 | 4373 |
| 1876 | Mary | 9.2% | 37.1% | 0.166 | 1875-1876 | 0.0029 | 0.0064 | 0.0064 | 8% | 0 | 3672 |
| 1877 | Mary | 8.6% | 37.4% | 0.150 | 1876-1877 | 0.0039 | 0.0102 | 0.0102 | 10% | 0 | 3917 |
| 1878 | Mary | 9.3% | 36.4% | 0.196 | 1877-1878 | 0.0055 | 0.0141 | 0.0141 | 10% | 0 | 4580 |
| 1879 | Mary | 8.4% | 34.7% | 0.144 | 1878-1879 | 0.0070 | 0.0175 | 0.0175 | 13% | 0 | 3480 |
| 1880 | Mary | 8.8% | 36.0% | 0.175 | 1879-1880 | 0.0052 | 0.0129 | 0.0129 | 11% | 0 | 5320 |
| ave. years 1870s | Mary | 9.6% | 38.2% | 0.178 | ave. y-to-y 1870s | 0.0042 | 0.0104 | 0.0180 | 10% | 0.6 | 3922 |
| grp 1870s | Mary | 9.1% | 37.1% | 0.165 | grp 1860s to 70s | 0.0219 | 0.0164 | 0.0174 | 18% | 1 | 38988 |
| | | | | | Name Communication Across the Nineteenth Century | | | | | | |
| 1800-9 to 1900-9 | | | | | | 0.4479 | 1.5132 | 2.0621 | 1753% | 5 | 82319 |

## Appendix C
## Mary, Group Polarization, and Symbolic Consensus

An extensive consensus about the value of Mary, as revealed in female names, emerged in England in the late sixteenth and early seventeenth centuries. In the early middle ages, Mary was considered a name too sacred to be applied to new uses. Few churches were dedicated to Mary in England through the early eighth century, and the first documented use of Mary as a personal name in England was in the eleventh century.[126] Table C1 shows the popularity of Mary in various name compilations over time. The popularity of the name Mary rose from less than 1% of females in the fourteenth century to about 20% of females from the late seventeenth century through the early nineteenth century (see also Table 3 in main text). The change appears to have been rather sudden and concentrated in the late sixteenth and early seventeenth centuries. The extent of this new consensus is astonishing: over the eighteenth and early nineteenth centuries, perhaps 30% of women who had at least one daughter had a daughter named Mary.[127]

The rise of this consensus contrasts sharply with contemporaneous political and religious polarization in England. In 1534 King Henry VIII broke from Catholic authority, established a new Church of England with himself as head, and in defiance of Roman judgement had his marriage to Anne Boleyn declared legal. All ecclesiastical and government officials were required to sign an oath of loyalty to the Church of England. The conflict between the Church of England and the Roman Catholic Church was largely political, not religious. But it occurred in the context of the Protestant Reformation, which provided detailed and passionate criticism of Roman Catholic doctrine and practices. Thus there was an environment with rich resources for creating enduring polarization between Protestants and Catholics.

---

[126] Withycombe (1945/1977) p. 211.
[127] For discussion of this statistic and details of its calculation, see Section II A.

| Table C1 Percent of Females Named Mary | | |
|---|---|---|
| Year, Location | Percent | Sample Size |
| 1210, South 3) | 0.0% | 173 |
| 1270, Rutland 4) | 0.0% | 206 |
| 1300, Lincoln 5) | 0.6% | 1213 |
| 1350, Hereford 9) | 0.7% | 576 |
| 1350, Yorkshire 9) | 0.2% | 1794 |
| 1560, Canterbury 17) | 7.3% | 661 |
| 1560 Gloucester 18) | 6.6% | c. 4000 |
| | | |
| 1530, North/Cumbria 13) | 1.9% | 852 |
| 1550, North/Cumbria 13) | 1.9% | 1491 |
| 1580, North/Cumbria 13) | 3.1% | 3750 |
| 1610, North/Cumbria 13) | 6.4% | 4000 |
| 1640, North/Cumbria 13) | 9.9% | 2888 |
| 1670, North/Cumbria 13) | 14.1% | 3813 |
| 1700, North/Cumbria 13) | 16.2% | 3064 |
| 1730, North/Cumbria 13) | 16.7% | 2038 |
| 1760, North/Cumbria 13) | 18.1% | 2830 |
| 1790, North/Cumbria 13) | 19.4% | 2139 |
| | | |
| 1620, Yorkshire 12) | 16.7% | 342 |
| 1670, Yorkshire 12) | 20.6% | 228 |
| 1720, Yorkshire 12) | 25.7% | 413 |
| 1770, Yorkshire 12) | 22.8% | 381 |
| | | |
| 1625, England 20) | 17.0% | n.a. |
| 1675, England 20) | 20.5% | n.a. |
| 1725, England 20) | 20.0% | n.a. |
| 1775, England 20) | 24.0% | n.a. |

Note: The number following the location indicates the data source. See References.

Protestants and Catholics historically have clashed sharply over Mary, the mother of Jesus. Catholics highly venerate Mary as the Mother of God, a model disciple of Christ, and Queen of Heaven. Repeatedly invoking Mary's name with the prayer Ave Maria ("Hail Mary") was a standard Catholic devotional practice from the fourteenth century. In 1572 Pope Gregory XIII proclaimed the Feast of the Most Holy Rosary, an official celebration of a prayer that repeatedly calls to Mary for her spiritual help.[128] The significance of Mary in Roman Catholicism is underscored in doctrines concerning Mary's immaculate conception and bodily assumption into heaven.

---

[128] For a history of the Rosary and other Marian prayers, see http://www.familyrosary.org/bk-hist.htm.

Protestants have tended to consider Catholic Marian practices and beliefs to be unbiblical, irrational, or idolatrous.[129] Here is a typical view from *The Weekly Observer* in 1716:

> *A Papist is an Idolater, who worships Images, Pictures, Stocks and Stones, the Works of Men's Hands; calls upon the Virgin* Mary *[distinctive typeface in original], Saints and Angels to pray for them...*[130]

The *Puritan Book of Discipline*, drafted about 1586 and subsequently widely discussed among Puritan leaders, declared:

> *Those that present unto baptism, ought to be persuaded not to give those that are baptized the names of God, or of Christ, or of angels, or of holy offices, as of Baptist or Evangelist, &c. nor such as savour of paganism or Popery....*[131]

It difficult to understand how the name Mary would not be an element in this category of symbols. Relative to the symbolic economy of Catholicism, Protestantism devalorizes Mary.

In the late sixteenth century the name Mary also had an important, polarizing political reference. As Queen of England from 1553-58, Mary I, the daughter of Henry VIII and Catherine of Aragon, tried to turn England back to Catholicism. She married Philip II of Spain, a Catholic king with little respect for the nascent English Parliament. She repealed Protestant legislation and burned about 300 persons as heretics. Under her reign England's financial and military strength waned amidst the religious and political turmoil. At least as reflected through historians' writings, Mary I, also known as "Bloody Mary," was widely feared and despised.

Subsequent events re-enforced tensions between representations of Catholicism and Anglicanism or Protestantism. Elizabeth I, who succeeded Mary I, supported the Church of England and Protestant forces in Scotland and continental Europe. She had Mary, Queen of Scots beheaded for conspiring with Catholics to overthrow her reign. In 1605 Guy Fawkes and four other Catholic radicals were caught attempting to blow up the House of Lords and kill King James I. November 5 thus became associated with "Guy Fawkes Day" celebrations, which often involved burning the Pope in effigy. Tensions continued with Charles I's marriage to a French Catholic princess (1625), the English Civil War and the establishment of a Commonwealth lead by the Puritan Oliver Cromwell (1649-58), the English colonization of largely Catholic Ireland, the Restoration and ascension of the Catholic James II (1685-88), and William of Orange and Mary II overthrowing James II and becoming co-sovereigns to protect Protestantism in England (1689).

---

[129] For discussion of Mary in Roman Catholicism, along with dialog with and links to Protestant criticism, see http://ic.net/~erasmus/ERASMUS9.HTM
[130] *The Weekly Observer*, June 1716, quoted in Haydon (1994) p. 22.
[131] The *Puritan Book of Discipline* is also known as *A Directory of Church-Government*. An English translation (the original was in Latin) can be found in Neal (1837), vol. III, App. No. IV. The quote is from p. 495. An important publication of this book occurred in 1644. For details on authorship, discussion, and influence of the book, see Pearson (1925) pp. 141, 257, 397-8. Cf. Wilson (1998) p. 193-4.

Religious and political polarization related to Catholicism occurred despite a relatively small number of Catholics in England. Catholics in England in 1680 and 1770 comprised less than 2% of the population; Anglicans and other Protestant denominations were by far numerically dominant.[132] Of course prior to 1534 all churches in England and most of the population were at least nominally Roman Catholic, and this historical legacy probably contributed to the perceived threat of a (re)turn to Catholicism. France and Spain, Catholic countries with a long history of wars with England, associated Catholicism with threats from foreign enemies. In addition, the Irish remained Catholic. Tensions associated with England's colonization of Ireland and cultural differences between the English and the Irish combined to re-enforce polarization along boundaries of religion.

Religious and political polarization had many concrete manifestations. Laws were passed that forbid Catholics to practice law, to serve as officers in the army or navy, to get university degrees, or to vote in local and parliamentary elections. Catholics were required to be married in an Anglican church, have their children baptized in an Anglican church, and be buried in an Anglican cemetery. There were widespread panics in 1715 and 1740 about an allegedly imminent Catholic uprising. In 1780, 60,000 persons gathered to march on Parliament to demand the repeal of the Catholic Relief Act of 1778, which relaxed some restrictions on Catholics. The march turned violent, and the ensuring riots, known as the Gordon Riots, resulted in 450 arrests and at least 285 deaths.[133]

Existing scholarship does not seem to have carefully examined the relationship between religious and political polarization and the use of the name Mary. Only one scholar seems to have considered the issue. Here is his analysis:

> ...[the name] Mary was in danger of becoming obsolete at the close of Elizabeth's reign, so hateful had it become to Englishmen, whether [Church of England] Churchmen or Presbyterians…. the fates came to the rescue of Mary, when the Prince of Orange landed at Torbay, and sate with James's daughter on England's throne. It has been a favorite ever since.[134]

The name Mary might have been hateful at the close of Elizabeth I's reign in reaction to the Catholic Queen Mary I (1553-1558), "Bloody Mary," and her successor, the Anglican Queen Elizabeth I (1558-1603), who was associated with a cultural and political golden era. The name Mary may have become popular when Mary II (James II's daughter) and William III (the Prince of Orange) became co-sovereigns in 1689 to prevent James II from attempting to turn England toward Catholicism. According to this account, Mary became a popular name when it became a representation for a Protestant queen who saved England from Catholicism.

The above account, however, is not consistent with the facts. As Table C1 shows, the name Mary had significant popularity at the close of Elizabeth I's reign (1603). Moreover, by the time Mary II assumed the throne (1689), about 20% of females were already being named Mary. The historical facts are painfully recalcitrant. In England from the late sixteenth century to the early nineteenth century, intense political and religious polarization coincided with the development of an astonishing degree of consensus about Mary as a personal name.

---

[132] Calculations based on Bossy (1975) pp. 185, 189, and Wrigley and Schofield (1981), Table A3.3.
[133] Haydon (1993) pp. 215, 237, and passim. For a broader historical analysis of anti-Catholicism, see Lockwood (2000).
[134] Bardsley (1880) p. 113.

This historical evidence points to the importance of analyzing carefully claims concerning the relationship between group polarization and the information economy. From a historical perspective, more research is needed to understand why such a new, extensive, and unlikely symbolic consensus emerged in England in the late sixteenth and early seventeenth centuries. With respect to current arguments connecting the Internet to political polarization, one should consider whether their persuasive logic is essentially similar to that of the above contrast between Catholic Queen Mary I and Protestant Queen Mary II.[135]

---

[135] See Sunstein (2001).

## I. Method of Analysis

All the samples of individual names were processed, to the extent possible, in the same way.  Where possible a sample included the given name, age, and sex.  From these fields new standardized name and standardized sex fields were constructed.  Given names were truncated to the shorter of 8 letters or the letters preceding the first period, space, hyphen, or other non-alphabetic character.  Use of upper or lower case letters was considered irrelevant.  Records with single letter names (abbreviations) were eliminated from the sample, as were records with generic names such as "Mr", "Mrs", "Widow", "Infant", etc.  Names were then standardized using the GINAP (version 1).[136]  The principle for the name standardization coding is to group together names that either sound the same, have the same public meaning, or changed only in the recording process (spelling errors, recording errors, etc.).  The standardization coding also corrects errors in the sex code for common, sex-unambiguous names.

Where available, the age field was used to construct samples based on birth year.   Thus the tables in the paper refer to birth years, i.e. the data relates to names given in the years indicated.   An average age for persons in the sample was estimated for samples that did not include individual age fields.  The average age for persons whose names were recorded at marriage was taken to be 25 years, which is consistent with estimates of the average age of marriage in the sixteenth through eighteenth centuries.[137]  Average age for a sample of adults not otherwise distinguished was taken to be 35 years.

In constructing communication statistics, in a few cases the sample or summary statistics available did not include a name that was one of the ten most popular names in the subsequent comparison year. The popularity of the name in the earlier year was assumed to be one-half of that of the least popular name observed or recorded from the earlier sample.

## II. US Name Data for Years 1801 to 1999

Tables 2, 4, and 7 and Chart 1 include statistics on US names from 1801 to 1999.  The underlying data are samples from the national censuses of 1850, 1880, and 1920, and from Social Security card applications 1910 to 1999.  Statistics for 1845 are averages of statistics for corresponding cohorts from the censuses of 1850 and 1880.  Statistics for 1875 are similarly averages of statistics for corresponding cohorts from the censuses of 1880 and 1920.  Statistics for 1915 are averages of statistics from the Census of 1920

---

[136] GINAP is available at http://users.erols.com/dgalbi/names/ginap.htm
[137] In England from 1550 to 1799, the age at first marriage for females was about 25 and for males about 27.  See Laslett et. al. (1980) Table 1.2, p. 21.

birth cohort and the Social Security applications for births 1910-19. Sample sizes (including average sample size for sample overlap years) are given in Table D1.

| year listed | years of birth | females | males |
|---|---|---|---|
| **Table D1** | | | |
| **US Name Statistics: Sample Sizes** | | | |
| 1805 | 1801-10 | 6,365 | 6,455 |
| 1815 | 1811-20 | 9,133 | 9,340 |
| 1825 | 1821-30 | 14,378 | 13,956 |
| 1835 | 1831-40 | 20,558 | 20,339 |
| 1845 | 1841-50 | 24,441 | 23,616 |
| 1855 | 1851-60 | 37,804 | 36,161 |
| 1865 | 1861-70 | 48,715 | 47,971 |
| 1875 | 1871-80 | 53,360 | 54,433 |
| 1885 | 1881-90 | 60,750 | 57,606 |
| 1895 | 1891-00 | 78,755 | 74,013 |
| 1905 | 1901-10 | 93,799 | 93,358 |
| 1915 | 1910-20 | 358,671 | 356,209 |
| 1925 | 1920-29 | 666,392 | 656,685 |
| 1935 | 1930-39 | 581,628 | 589,315 |
| 1945 | 1940-49 | 765,997 | 805,466 |
| 1955 | 1950-59 | 1,003,004 | 1,056,854 |
| 1965 | 1960-69 | 969,700 | 1,003,067 |
| 1975 | 1970-79 | 843,550 | 872,355 |
| 1985 | 1980-89 | 938,306 | 973,698 |
| 1995 | 1990-99 | 882,469 | 909,288 |

The samples from the censuses of 1850, 1880, and 1920 are 1-in-100 random samples of individuals, families, and dwellings. They thus include names grouped by family and dwelling. The 1850 sample does not include slaves. Persons not born in the US were excluded from the sample in order to focus on naming patterns in the US. Census samples after 1920 do not include personal names. For further documentation and the data, see Ruggles and Sobek (1997).

The Social Security name summaries are available for 1900 to 1999. They are calculated from 5% samples of Social Security card applications for persons born on US soil. The data is presented as counts for the top 1000 names by decade and by sex for persons born in those decades. For further documentation and the data, see Shackleford (2000).

### III. UK Name Data for Years 1800 to 1994

Tables 2, 3, 5, 6, and A1 include statistics on UK names from 1800 to 1994. Statistics for 1800-1880 have been constructed from age cohorts of the Census of 1881. For documentation and the data, see GSU-FFHS (1997). This source provides a complete enumeration of all Census of 1881 records for England and Wales. Persons not born in England and Wales were eliminated in order to focus on naming in England and Wales. The data include location codes, and thus statistics can be calculated for specific geographic areas (see year 1825 data for some locations in Tables 5 and A1). Table D2 gives the sample sizes for statistics in Tables 3 and 6.

| Table D2 UK Name Statistics 1800-1880: Sample Sizes | | | |
|---|---|---|---|
| year listed | years of birth | females | males |
| 1800 | 1800 | 9,190 | 7,331 |
| 1810 | 1810 | 27,988 | 24,265 |
| 1820 | 1820 | 55,358 | 50,036 |
| 1830 | 1830 | 68,941 | 62,357 |
| 1840 | 1840 | 99,828 | 94,183 |
| 1850 | 1850 | 129,977 | 119,392 |
| 1860 | 1860 | 199,693 | 185,195 |
| 1870 | 1870 | 232,213 | 229,511 |
| 1880 | 1880 | 282,986 | 281,717 |

In Table 3, data for 1900 are from Dunkling (1977) and data for 1925 are from Dunkling (1995). In Table 6, middle set of figures, for years 1850, 1875, and 1900, data are from Dunkling (1977), and for years 1925, 1950, and 1975 data are from Dunkling (1995). Dunkling (1977) and Dunkling (1995) do not provide sample sizes and present slightly different name popularity counts for years listed in both.

Data for birth years 1944 for 1994 are from Merry (1995). The names are from the National Health Service Central Register and pertain to persons born in the year indicated. Table D3 gives sample sizes.

| Table D3 UK Name Statistics 1944-1994: Sample Sizes | | | |
|---|---|---|---|
| year listed | years of birth | females | males |
| 1944 | 1944 | 373,377 | 382,217 |
| 1954 | 1954 | 387,138 | 394,627 |
| 1964 | 1964 | 502,850 | 504,911 |
| 1974 | 1974 | 350,305 | 357,274 |
| 1984 | 1984 | 331,682 | 347,467 |
| 1994 | 1994 | 329,739 | 347,986 |

Data for 1805, Manchester in Table 5 are from Galbi (source 11). These data represent factory workers whose names were recorded in a 1818-19 House of Lords investigation of factory conditions.

## IV. Names in England before 1800

Tables 5, 8 and A1 include statistics on names in England before 1800. The numbers followed by a right parenthesis in Table 5 are source numbers. The primary data source section of References provides complete citations for each source as well as the associated source number. The locations listed in Tables 8 and A1 correspond to the locations listed in Table 5. The associated sources are the same as those indicated in Table 5. The years associated with the locations are the estimated average year of birth for the persons in the sample. For most persons names were probably given within plus or minus twenty years of the listed year.

Table 8 gives name communication statistics for the century following the Norman Conquest. These statistics depend on an estimate of the popularity distribution of the ten most popular names in 1166 and an estimate of the popularity distribution of those names in 1066. Based on data in Table 5, a log-linear popularity distribution for 1166 was estimated with top name popularity of 10% and top ten name popularity of 45%. Based on Barlow *et. al.* (1976) Table 7 p. 185, which shows Continental Germanic and Biblical, Greek, and Latin names having a total popularity of 10.6% in 1066, the popularity in 1066 of the top ten names in 1166 was chosen as 10% of their 1166 popularity, i.e. 4.5%. In addition, the information $I_s$ in the top ten names was taken to be 0.4, a reasonable "equilibrium" value about 1300 (see Table 5). Using these parameters a log linear distribution was estimated for the popularity in 1066 of the top ten names in 1166. The name communication statistics for the 1066-1166 Conquest Model are based on these estimates.

For the sixteenth through the eighteen centuries, parish registers provide a large amount of name data. These registers have been studied as part of important and extensive work on family and population history. See Wrigley and Schofield (1981). As far as I am aware, however, these efforts have not encompassed personal names, nor have they produced datasets useful and available for such study.

Members of the Society for Creative Anachronism (SCA) have been a leading force in scholarship on names used before circa 1600. The Academy of Saint Gabriel provides an institutional focus for SCA name research.[138] The Medieval Names Archive, published by Arval Benicoeur (Joshua Mittleman), provides an extensive collection of sources for studying names in many different languages and places around the world.[139] SCA participants typically have two names. In this paper, sources are listed under a SCA participant's current medieval name when it appears to provide the primary authorial

---

[138] The Academy of Saint Gabriel is online at http://www.s-gabriel.org/
[139] The the Medieval Names Archive online at http://www.panix.com/~mittle/names/

identifier for the work.  The current name is given in parentheses following the medieval name.

Genealogical researchers are also rapidly expanding historical knowledge about names. The Internet is a powerful tool for genealogical researchers to network and share information.  Nonetheless, effective scholarly standards for sharing research are still only developing.  Such standards help researchers with different perspectives and different questions to aggregate and analyze knowledge independently.  Compilations of names from genealogical research potentially could be an important source for studying name history and information economies.